\documentclass[11pt,preprint]{aastex}

\def\asca{{\it ASCA}}

\def\pcmsq{{$\rm cm^{-2}$}}
\def\chisq{{$\chi^{2}$}}

\def\nh{{$N_{\rm H}$}}

\begin{document}

\title {SHEEP: THE SEARCH FOR THE HIGH ENERGY EXTRAGALACTIC POPULATION}

\author {K. Nandra\altaffilmark{1,2}, I. Georgantopoulos\altaffilmark{3},
A. Ptak\altaffilmark{4,5},T.J. Turner\altaffilmark{1,6} }

\altaffiltext{1}{Laboratory for High Energy Astrophysics, Code 660, 
	NASA/Goddard Space Flight Center,
  	Greenbelt, MD 20771}
\altaffiltext{2}{Universities Space Research Association}
\altaffiltext{3}{National Observatory of Athens, Greece}
\altaffiltext{4}{Department of Physics, Carnegie Mellon University, 5000 Forbes Avenue, Pittsburgh, PA 15213}
\altaffiltext{5}{Department of Physics and Astronomy, Johns Hopkins University, 3400 North Charles Street, Baltimore, MD 21218}
\altaffiltext{6}{University of Maryland, Baltimore County, 
1000 Hilltop Circle, Baltimore, MD 21250}

\slugcomment{Submitted for publication in
{\em The Astrophysical Journal}}

\begin{abstract}

We present the SHEEP survey for serendipitously-detected hard X-ray
sources in ASCA GIS images. In a survey area of $\sim 40$~deg$^{2}$,
69 sources were detected in the 5-10 keV band to a limiting flux of
$\sim 10^{-13}$~erg cm$^{-2}$ s$^{-1}$. The number counts agree with
those obtained by the similar BeppoSAX HELLAS survey, and both are in
close agreement with ASCA and BeppoSAX 2-10 keV surveys. Spectral
analysis of the SHEEP sample reveals that the 2-10 and 5-10 keV
surveys do not sample the same populations, however, as we find
considerably harder spectra, with an average $\Gamma\sim1.0$ assuming
no absorption. The implication is that the agreement in the number
counts is coincidental, with the 5-10 keV surveys gaining
approximately as many hard sources as they lose soft ones, when
compared to the 2-10 keV surveys. This is hard to reconcile with
standard AGN ``population synthesis'' models for the X-ray background,
which posit the existence of a large population of absorbed
sources. We find no evidence of the population hardening at faint
fluxes, with the exception that the few very brightest objects are
anomalously soft.  53 of the SHEEP sources have been covered by ROSAT
in the pointed phase. Of these 32 were detected. An additional 3 were
detected in the RASS. As expected the sources detected with ROSAT are
systematically softer than those detected with \asca\ alone, and of
the sample as a whole. Although they represent a biased subsample, the
ROSAT positions allow relatively secure catalog identifications to be
made. We find associations with a wide variety of AGN and a few
clusters and groups. At least two X-ray sources identified with high-z
QSOs present very hard X-ray spectra indicative of absorption, despite
the presence of broad optical lines. A possible explanation for this
is that we are seeing relatively dust-free ``warm absorbers'' in high
luminosity/redshift objects.  Color analysis indeed indicates that
many of the spectra are not consistent with a simple, absorbed power
law. The spectra are likely to be complex, with an absorbed hard power
law and scattered or ``leaky'' component in the soft X-rays. Many are
also consistent with a reflection dominated spectrum. Our analysis
defines a new, hard X-ray selected sample of objects - mostly active
galactic nuclei - which is less prone to bias due to obscuration than
previous optical or soft X-ray samples. They are therefore more
representative of the population of AGN in the universe in general,
and the SHEEP survey should produce bright examples of the sources
that make up the hard X-ray background, the majority of which has
recently been resolved by Chandra. This should help elucidate the
nature of the new populations.

\end{abstract}


\section{INTRODUCTION}
\label{Sec:Introduction}

ROSAT observations have shown that the diffuse X--ray background (XRB)
in the soft X-ray band (0.5-2 keV) is made up of discrete
sources, primarily standard, broad-line QSOs (Shanks et al. 1991;
Hasinger et al. 1998; Schmidt et al. 1998).  It is puzzling, however,
that the X-ray spectra of AGN above $\sim 2$ keV, which typically show
an intrinsic power law index of $\sim 1.9$ (Nandra \& Pounds 1994), is
so much steeper than the observed background in the same band, with
$\Gamma\sim 1.4$ (Marshall et al. 1980). This ``spectral paradox'' has
received much attention, and a consensus appears to be emerging as to
its resolution.  Setti \& Woltjer (1989) suggested that the spectral
paradox can be solved by hypothesizing that large numbers of AGN are
heavily absorbed in the X-ray band, consistent with Seyfert
unification schemes (Lawrence \& Elvis 1982; Antonucci \& Miller
1985). Comastri et al. (1995), and other authors (e.g. Madau,
Ghisselini \& Fabian 1994; Gilli et al. 1999, 2001) have shown that
AGN spectra with a range of absorbing column densities can indeed be
made to fit the XRB spectrum consistently with the number counts.
Such models of the XRB are the most promising to date, and agree with
many observables, including the range of spectra seen in nearby,
bright AGN. They predict that the major contributors to the XRB are a
large population of highly-absorbed AGN at moderate-high
redshift. This population of objects has never been directly
observed. This is perhaps not surprising, given that traditional
UV-excess and soft X-ray surveys - which have detected most of the AGN
we know of so far - are biased against strongly-absorbed objects.

The best method of uncovering obscured AGN is in the hard X-ray band,
and a large population of such sources is implied. The ROSAT 0.5-2 keV
number counts can be converted into 2-10 keV counts by extrapolating
the ROSAT flux assuming the mean spectrum of the ROSAT sources of
$\Gamma=2$. This exercise under-predicts the number counts observed by
\asca\ by a factor $\sim 2$ (Georgantopoulos et al. 1997; Cagnoni et
al. 1998; Ueda et al. 1999a). This immediately implies the presence of
a large population with flat or absorbed spectra.  Optical
spectroscopic identifications of ASCA sources have identified a few
examples of obscured AGN at high redshift. For example, Boyle et
al. (1998) have reported the discovery of an X-ray obscured quasar at
z=0.67 and Georgantopoulos et al. (1999) have described the properties
of an even more extreme z=2.35 quasar (see Akiyama, Ueda \& Ohta 2002
for another example). Chandra has now also uncovered a few examples of
type II QSOs (Norman et al. 2001; Stern et al. 2002).  The inferred
column densities for these sources are $>10^{23}$~cm$^{-2}$ which
makes them extremely difficult to detect or identify in soft X-ray
surveys and, if the absorbing material is dusty, heavily redenned and
weak in the optical.  There may be a vast population of such hidden
beasts lurking in the hard X-ray sky.

A promising start towards uncovering this population has been made
using the BeppoSAX HELLAS survey. Fiore et al. (1999) presented the
first results of a survey in the 5-10 keV with the BeppoSAX MECS
instrument. Their survey covered an area of 50 deg$^{2}$ and detected
$\sim$150 sources above a limiting flux of about $5\times
10^{-14}$~erg cm$^{-2}$ s$^{-1}$. The updated HELLAS survey of Fiore
et al. (2001) covers a larger area of 85 deg$^{-2}$, with 147 sources.
The number count distribution, logN-logS, presents a Euclidean slope
with $\gamma=1.56\pm0.14$, in apparent agreement with the population
synthesis models (Comastri et al. 2001). At the survey's limiting flux
$\sim30$ per cent of the 5-10 keV XRB has been resolved.  Catalog
cross-correlations and the results of an optical ID program showed a
high proportion of AGN, many of which are heavily obscured and some of
which are at moderately high redshifts (Fiore et al. 1999; Fiore et
al. 2001).  This survey contains several examples of obscured AGN both
because of its large area and of the very hard X-ray band
employed. Preliminary results from {\it XMM} on the Lockman hole field
(Hasinger et al. 2001) has extended the logN-logS a factor of twenty
deeper (see also Baldi et al. 2001). It appears to present a Euclidean
slope all the way down to the limiting {\it XMM} flux of
$2.4\times10^{-15}$. At this flux limit 60 per cent of the 5-10 keV
XRB has been resolved.

New data from Chandra have now provided a breakthrough in our
understanding of the XRB, but have also raised further
questions. Mushotzky et al.  (2000) presented observations of a deep
Chandra-ACIS field.  They detected a few tens of sources most of which
have been spectroscopically followed up with the Keck telescope
(Barger et al. 2001).  Surprisingly there were no numerous examples of
the long sought obscured ``type II QSO'' population (note we adopt the
traditional definition of type I and type II objects based on their
optical, rather than X-ray properties). Instead two other distinct
populations appear.  One is of bright early-type galaxies.  These
galaxies are ``passive'', in that they present no clear sign of AGN
activity in their optical spectra. The other population of X-ray
sources is of hard sources with very faint or non-existent optical
counterparts (even in deep Keck images, implying $B>28$ in some
cases). This immediately highlights a problem with the Chandra
population - that they are too faint for effective optical or X-ray
followup.  Deeper observations in the Chandra Deep Field South (CDFS),
and in the Hubble Deep Field North (HDFN) respectively confirm the
above findings (Tozzi et al. 2001; Brandt et al.  2001a) .  For
example, Brandt et al. (2001a) detect 12 sources in the area of the
HDF (at a flux limit of $\sim2\times10^{-16}$ $\rm erg~cm^{-2}~s^{-1}$
in the 2-8 keV band) of which 4 are passive early-type galaxies while
only 3 are broad-line AGN. The early, tentative suggestion is that the
bulk of the X--ray background may arise from relatively low
luminosity, low redshift AGN of uncertain character.

Further progress is possible by undertaking a large area hard X-ray
survey, which can in principle reveal bright examples of the faint
Chandra sources, to help understand their nature. HELLAS has already
made some progress in this regard. Here we present a similar
survey with the \asca\ GIS instruments, called SHEEP (Search for the
High Energy Extragalactic Population).

\section{DATA ANALYSIS}

The Gas Imaging Spectrometers (GIS; Tashiro et al. 1995) aboard the
\asca\ satellite (Tanaka, Inoue \& Holt 1994) are ideal instruments
for undertaking a wide angle hard X-ray survey.  They have a
relatively large field-of-view ($\sim 0.3$ deg$^2$) and cover a broad
energy range (0.7-10 keV) with good sensitivity, low background and
moderate spatial resolution (HPD$\sim 3$~arcmin; Serlemitsos et al.
1995). A few dedicated survey projects were undertaken during the
mission, but as \asca\ was in operation for over 6 years mainly in
pointed mode, it thereby accumulated serendipitous data on a large
number of fields. We have used as a starting point for the SHEEP
survey the ``Tartarus'' database of \asca\ observations of AGN (Turner
et al. 2002). This is ideal for our purposes, as most of the original
ASCA targets are point-like, and many of them relatively weak. Fields
with bright or extended targets are less useful, as they spread out
over the GIS field of view, making it difficult to detect objects
serendipitously.

\subsection{Field Selection}

We began with 464 ASCA observation sequences in version 1 of the
Tartarus database.  We rejected fields with a) exposure less than 30
ksec (combined GIS2+GIS3) b) Galactic latitude $|b|<20$ and c) targets
brighter than 0.02 ct s$^{-1}$ in the 5-10 keV band. Bright targets
can dominate the whole field due to the extended wings of the PSF,
dramatically increasing the effective background. We also rejected a
few fields for miscellaneous reasons, for example if the pointings
were at extended sources (another was rejected due to contamination by
a bright off-axis source). Finally, where multiple observations of the
same target source existed, we chose the one with the highest
exposure. This gave a total of 149 fields used in our survey. We
restricted our analysis to the central 18 arcmin region of the GIS
detectors and, as described below, excluded an area of 5 arcmin radius
around the ASCA target. This results in a total area for the survey of
38.9~$\rm deg^2$.

\subsection{Data reduction}

Cleaned event files were taken from the Tartarus processing described
in Turner et al. (2002).  We extracted sky images in the 5-10 keV band
for the GIS2 and GIS3 detectors from these using FTOOLS V5.0 and
co-added them to produce a combined image. We then smoothed this with
a gaussian of FWHM 1.8 arcmin and examined it visually for sources,
marking the position of all candidates, regardless of whether they
were close to the target position or not. We extracted counts from a
cell of radius 5 pixels (75 arcsec). The area of the resultant
detection cell was approximately $4$ arcmin$^2$. We determined the
expected background in the detection cell using the ``mkgisbgd'' tool,
which averages a large number of deep GIS pointings with sources
removed.  The predicted background in the cells was on average $\sim
17$~ct, but could be as low as $\sim 7$~ct. This means that a Gaussian
approximation to the Poisson statistics could not be employed. We
therefore determined the Poisson significance of each source based on
the predicted and actual counts. We also performed spot checks on some
images using a Mexican hat wavelet transform, and found our detection
mechanism to be robust. After source detection, we discarded all
objects with a position within 5 arcmin of the intended target of the
observation. We chose this relatively large exclusion radius not only
to exclude the original ASCA targets, but to account for possible
extent in the targets and spurious sources representing the PSF wings
(the ASCA PSF is highly azimuthally asymmetric; Serlemitsos et
al. 1995).

\section{THE SURVEY}

\subsection{Detected Sources}

Table~\ref{tab:sample} shows the objects detected in our survey.  We
detected 69 sources with a significance above a Poisson probability
threshold of $3\times 10^{-5}$ or approximately 4.5$\sigma$ for the
equivalent Gaussian distribution. Considering our survey area and
detection cell size this results in less than one source expected by
chance. The table shows the (centroided) source position, the Poisson
probability and the equivalent gaussian ``sigma''. The count rate
shown is the raw rate summed over the two GIS detectors in the 5-10
keV band. In order to convert these into a flux, we need to account
for the size of the detection cell and the instrumental sensitivity at
the particular off-axis angle of the source. We achieved this by
extracting the spectrum of the source, and using the ``ascaarf'' task
to determine the effective area for the extracted region. This was
then compared to the on-axis effective area and the count rate
corrected to represent an effective on-axis rate, also shown in the
Table. Count rates were converted to fluxes assuming a spectrum of
$\Gamma=1.6$, giving a conversion factor of $1.24 \times 10^{-10}$~erg
cm$^{-2}$ s$^{-1}$ per combined GIS ct s$^{-1}$. The assumed spectrum
is in fact rather softer than our inferred mean spectrum ($\Gamma\sim
1.0$), but we adopt this value for the Table as it was also used by
HELLAS (Fiore et al. 2001), and the ASCA 2-10 keV survey of Cagnoni et
al. (1998) and Della Ceca et al. (1999), allowing an easier
comparison. Adoption of the flatter slope makes a difference of 10~per
cent in the derived flux, so this difference does not introduce a
severe error, but we discuss the effects of the assumed slope on our
results below. Our faintest source has a corrected count rate of $7
\times 10^{-4}$~ct s$^{-1}$ corresponding to a flux of $9 \times
10^{-14}$ erg cm$^{-2}$ s$^{-1}$ (5-10 keV) with our assumed spectrum.

\subsection{Number Counts}

\begin{figure}[t]
\plottwo{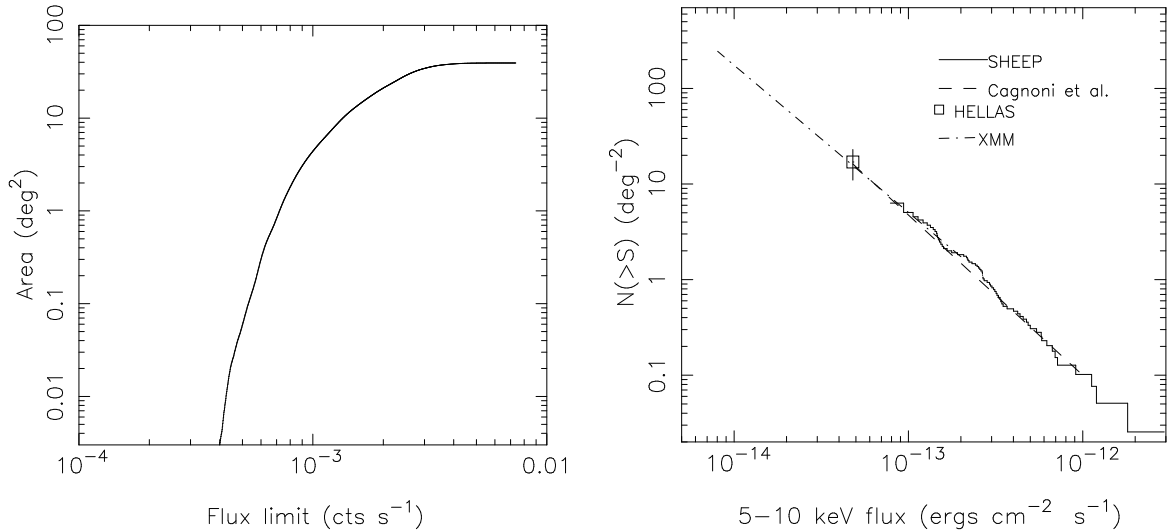}{fig_xnc.ps}
\caption{(left panel) The area covered by our survey as a function of
the flux limit. (right panel) The 5-10 keV SHEEP logN-logS (solid
line) compared to the deepest point in the BeppoSAX HELLAS survey
(open square; Fiore et al. 2001) and the ASCA 2-10 keV logN-logS
(dashed line; Cagnoni et al. 1998) converted to the 5-10 keV band
using $\Gamma=1.6$, The dash-dot line represents the XMM logN-logS
derived by Baldi et al. (2001).
\label{fig:xnc}}
\end{figure}

In Fig.~\ref{fig:xnc} we plot the number count distribution,
logN-logS, in the 5-10 keV band for our sample. To calculate this, it
is first necessary to determine the detection threshold as a function
of area for each image (left panel of Fig.~\ref{fig:xnc}).  We did
this by producing the exposure and effective area maps using
the ``ascaexpo'' and ``ascaeffmap'' tasks. A background map was
constructed using the ``mkgisbgd'' tool, and upper limits calculated
assuming Poisson statistics. This was then converted into an
effective flux limit using the effective area map. We show the
resultant integral logN-logS distribution in the right panel of
Fig.~\ref{fig:xnc}.  A maximum likelihood fit gives $N(>S) = 5
(S/S_{0})^{-1.68^{+0.25}_{-0.25}}$~deg$^{-2}$, where $S_{0}=
10^{-13}$~erg cm$^{-2}$ s$^{-1}$.  We find excellent agreement with
the logN-logS derived from the BeppoSAX HELLAS survey (Fiore et
al. 2001), and the XMM number counts from Baldi et al. (2001), which
reach fainter fluxes. We can also compare with the ASCA 2-10 keV
logN-logS (Cagnoni al. 1998) converted to the 5-10 keV band again
assuming a mean source spectrum of $\Gamma=1.6$.  We see that the
Cagnoni et al. logN-logS, $N(>S)\approx 10^{-21} S^{-1.67}$~deg$^{-2}$
(in the 5-10 keV band) provides a good fit to both the SHEEP and
HELLAS data.  At the faintest flux probed by our survey we resolve
about 15\% of the 5-10 keV XRB, using the HEAO-1 XRB normalization of
Marshall et al. (1980). We note again here that in calculating the
fluxes of the SHEEP sources, we have adopted $\Gamma=1.6$, softer than
the inferred mean for our sample. Using the most extreme mean spectrum
for our analysis of hardness ratios (see below) of $\Gamma=0.8$ would
lead to a 10~per cent increase in all the fluxes, and therefore a
$\sim 15$~per cent difference in logN-logS between SHEEP and the other
surveys.

\section{ROSAT observations and identifications}

\begin{figure}
\plotone{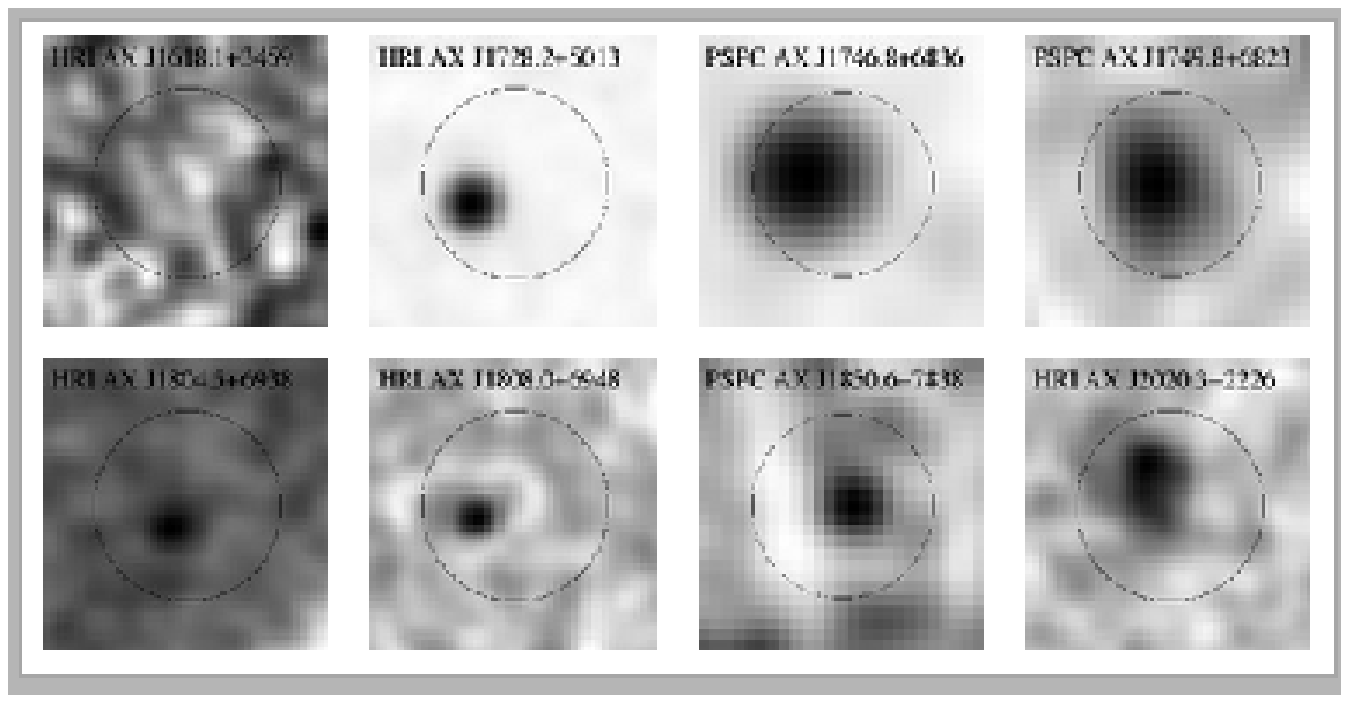}
\caption{ROSAT ``postage stamp'' images for the SHEEP sources. The circles
show a 2 arcmin radius centered on the ASCA position. 
\label{fig:rosat}}
\end{figure}

ROSAT observed the whole sky in the all-sky survey phase, and a very
large fraction of the SHEEP objects in its pointed phase (see
Table~\ref{tab:rosat} for details). Any available ROSAT data are
useful for our purposes, as they give both additional constraints on
the colors/spectrum of the SHEEP sources, and also better positional
accuracy for optical followup. We have cross-correlated the SHEEP
catalog with several ROSAT catalogs, specifically the Rosat All Sky
Survey (RASS) bright and faint-source catalogs, the WGACAT pointed
catalog, and the ROSHRI HRI pointed image catalog.  We found a total
of 35 SHEEP sources associated with a ROSAT source within 2 arcmin
($\sim2 \sigma$ GIS position), 3 of which were in the RASS catalogs
only. Some are detected in several catalogs and in some cases there
are several ROSAT catalog objects listed within 2 arcmin of the ASCA
position. In practice many of these may in fact be the same source, with
multiple listings in the catalogs. In all such cases we have examined
the ROSAT images (see Fig.~\ref{fig:rosat}) and determined whether
there are indeed multiple ROSAT counterparts within the ASCA error
box. We find no such case.

We have assessed the chance co-incidence of the associations between
the detected ASCA and ROSAT sources by offseting the SHEEP positions
by a few arc minutes, and repeating the cross-correlation.  Performing
10 such simulations we find that the number of expected false
coincidences are 0.8, 3.1 and 0.5 for the RASS, WGACAT and ROSHRI
respectively. Clearly most of the RASS and HRI detected objects are
secure, but there is some ($\sim 10$~per cent) probability that a
given WGACAT source is not in fact associated with the SHEEP source.

\begin{figure}[t]
\plotone{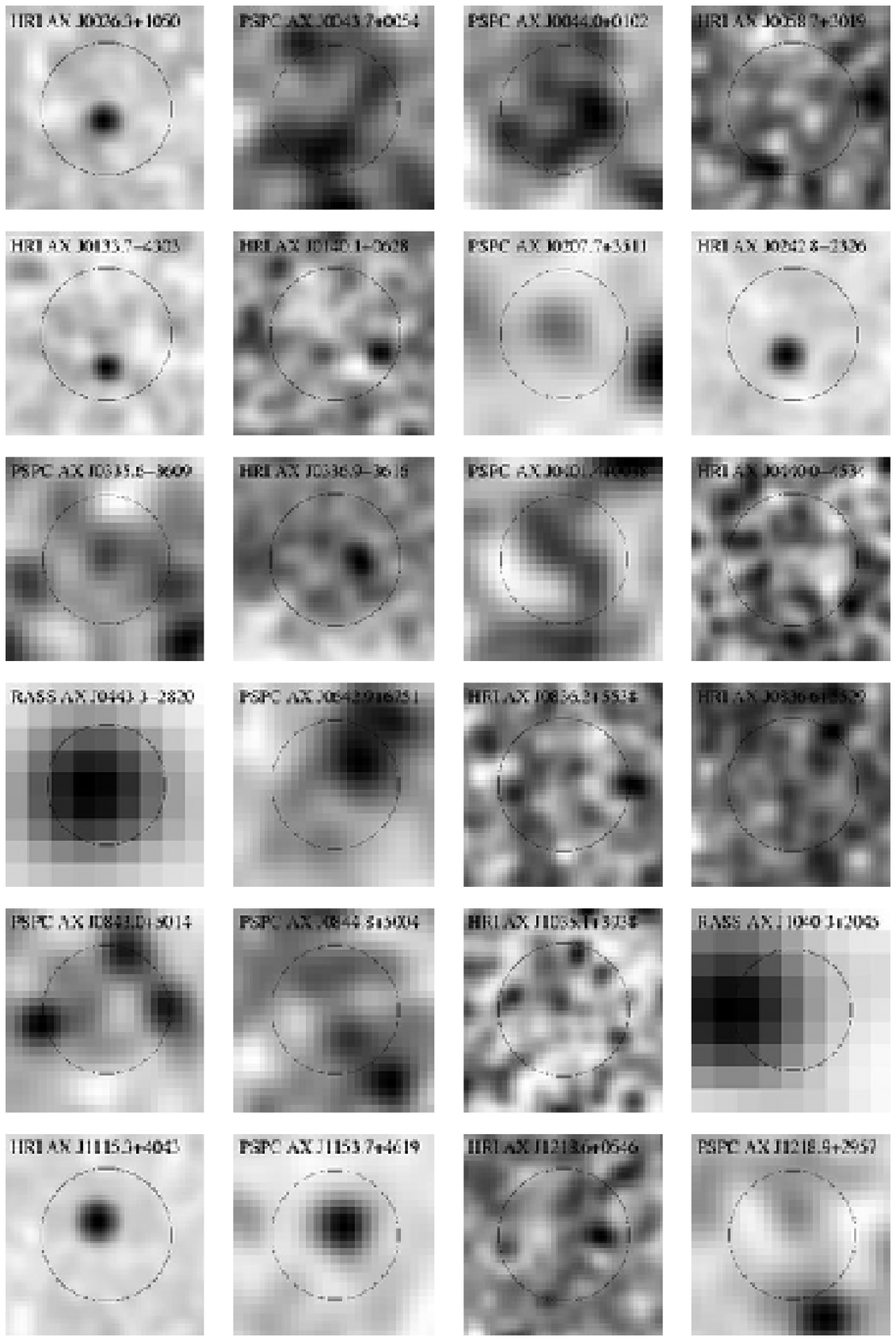}
\end{figure}

\begin{figure}
\plotone{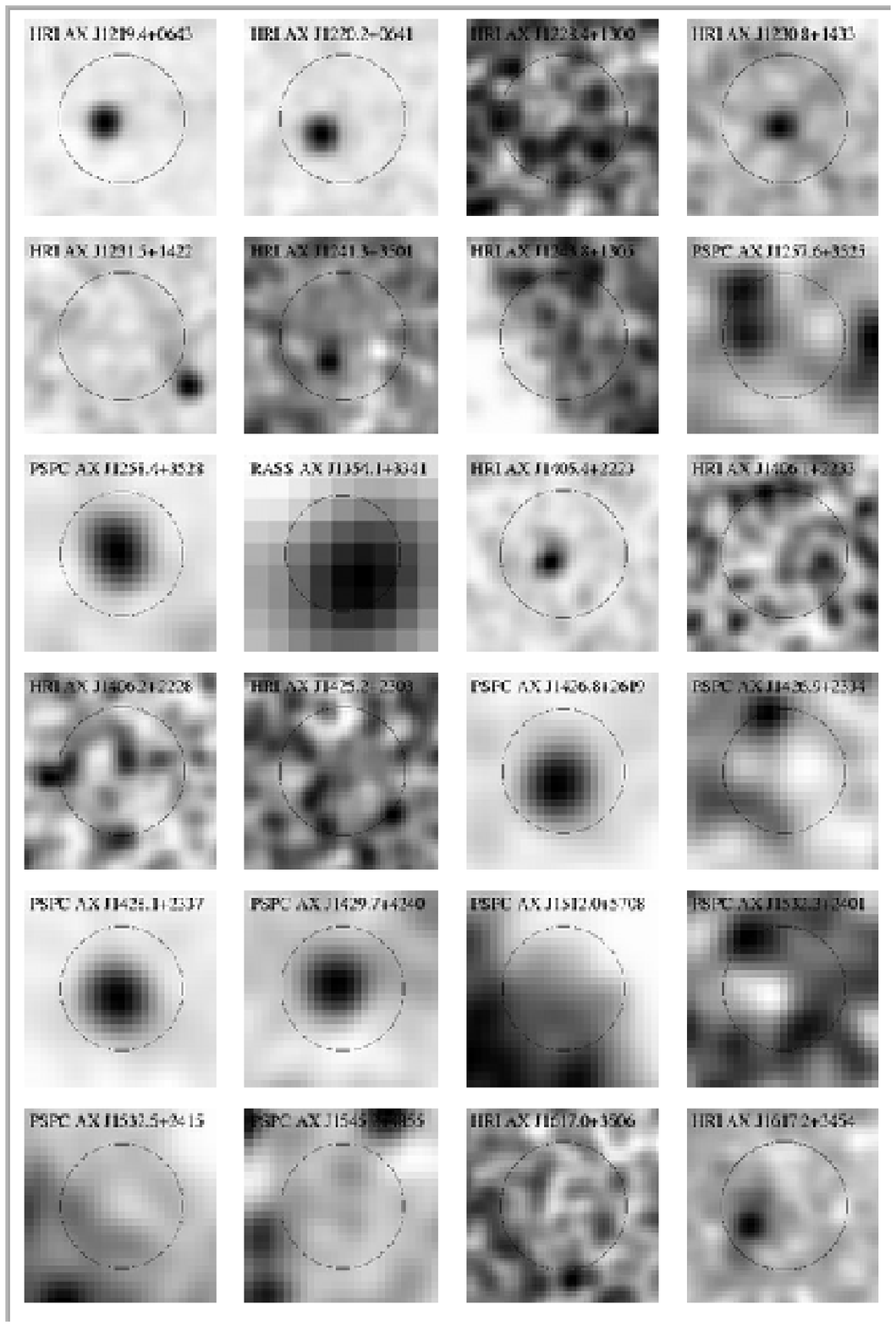}
\end{figure}

We have also independently analyzed the ROSAT data to determine upper
limits for the SHEEP sources which were observed by ROSAT in the
pointed phase, but not detected. In practice 53 SHEEP sources were
within the 2-degree diameter FOV of a pointed ROSAT PSPC observation,
and/or the ROSAT HRI FOV. Table~2 gives the details of these
observations, in which we preferentially quote values from the HRI
if possible, followed by pointed PSPC observations and finally RASS
data where no pointed observation exists. We have extracted images for
all of these, which are shown in Fig.~\ref{fig:rosat}, which also
shows images for the 3 additional SHEEP listed in the RASS bright and
faint-source catalogs. This independent analysis confirms the
catalog cross-correlations listed above, revealing 32 ROSAT detections
in pointed observations.  In the 21 cases where no source was
detected, we have derived upper limits to the ROSAT flux by extracting
the counts from the entire 2 arcmin GIS error box, renormalizing to
the 95~per cent PSF of the ROSAT PSPC or HRI, and calculating the
3$\sigma$ upper limit. The ROSAT non-detections cover a similar range
of exposure time and off-axis angles to the detections, implying that
the non-detections are not simply due to sensitivity
issues. Apparently therefore a large fraction of the SHEEP sources
($\sim 40$~per cent) have extremely hard spectra rendering them
undetectable by ROSAT. We discuss their spectral properties further
below. A less likely alternative is that they are highly variable.
The fraction of the SHEEP sources detected by ROSAT ($\sim 60$~per
cent) is remarkably similar to the equivalent fraction of
ROSAT-detected HELLAS sources (Vignali et al. 2001).

\section{Catalog Identifications}

The crucial remaining question for the X-ray background - which our
survey can help to answer - is that of the nature of the sources which
constitute the bulk of the XRB at hard X-ray energies. Our optical
follow-up work is just beginning, but some indications can be gleaned
without the use of telescope time by cross-correlating our X-ray
catalog with existing data. This job is considerably easier
for sources that have the more accurate ROSAT PSPC or (better still)
HRI positions, which reduces the possibility of chance coincidences. 

We have correlated the positions of the sources with ROSAT
counterparts with the NASA/IPAC Extragalactic Database (NED).  The
ROSAT/NED associations are shown in Table~\ref{tab:catids}.  19 SHEEP
sources have NED counterparts within 1 arcmin (PSPC position) or 15
arcsec (HRI position). Although the latter is rather larger than the
nominal positional error of the HRI, we allow for a large error in the
HRI catalog positions as the sources may be far off axis, and the
attitude solution is rather uncertain. Indeed we do find some very
secure counterparts to HRI sources (e.g. bright QSOs or Sy 1) that are
offset from the ROSAT positions by fairly large angles (see column 7
of Table~\ref{tab:catids}). The dominant population is clearly AGN,
but with many subclasses including classical QSOs, Seyfert 1s and more
obscured Seyferts. Our highest redshift source is the QSO FBQS
J125829.6+35284 at z=1.92. Unusually for a QSO this shows X-ray colors
indicating a very hard spectrum (Table~\ref{tab:sample}; see next
section). Another example is CRSS J1429.7+4240. These are unusual
because their optical spectral type indicates that there is an
unobscured view of the nucleus but the X-ray colors are indicative of
absorption.  We shall return to this in the discussion.

Despite these few sources with hard spectra, we shall show in the next
section that the ROSAT-detected sources are not a representative
subsample of the SHEEP sources as a whole. Indeed, as might be
expected, they are systematically softer. It is therefore extremely
important to find and identify the optical counterparts of the X--ray
sources that have not been detected by ROSAT, as these are the ones
most likely to be fruitful in determining the origin of the hard X-ray
background.  We have therefore cross-correlated the entire SHEEP
catalog with NED. Due to the large positional error of the GIS ($\sim
2$ arcmin), most of the resulting associations are probably random,
the bulk being with anonymous radio (NVSS: Condon et al. 1998; FIRST:
Becker, White \& Helfand 1995 ) or 2MASS (Cutri et al. 2000)
sources. These catalogs have rather high source densities and chance
associations are likely. One such association that is likely to be
real, however, is that of AX J1531.9+2420 with a radio loud quasar
discovered in the FIRST survey, FBQS J153159.1+24204 at z=0.631 (White
et al. 2000).  For the others, it will be rather difficult to obtain
unambiguous optical counterparts for the sources without improved hard
X-ray positions.

\section{Spectral properties}

\subsection{Hardness Ratios}

\begin{figure}[t]
\epsscale{0.8}
\plotone{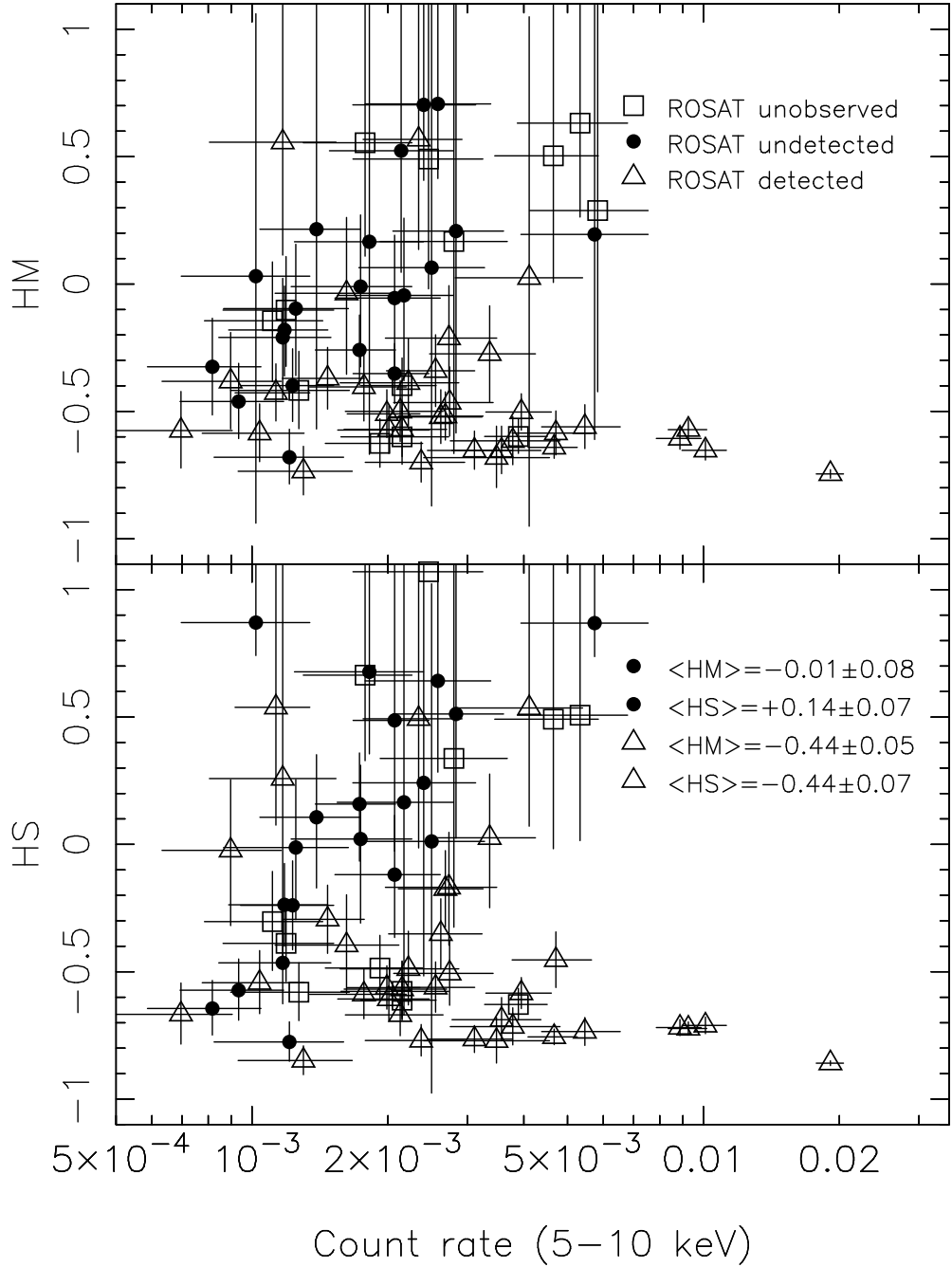}
\caption{Hardness ratios versus the 5-10 keV corrected count rate. HM
denotes the 5-10 vs. 2-5 keV hardness ratio and HS the 5-10 versus
0.7-2 keV hardness ratio.  Open squares are the 13 objects unobserved
by ROSAT in the pointed phase.  Filled circles were observed but not
detected (21 objects). The 35 sources observed and detected by ROSAT
are shown as the open triangles.  As expected, the ROSAT-detected
objects are significantly softer than those observed and not
detected. Additionally, they are significantly softer than the mean of
the sample as a whole, and form a biased subsample.
\label{fig:hr1}}
\end{figure}

\begin{figure}
\plotone{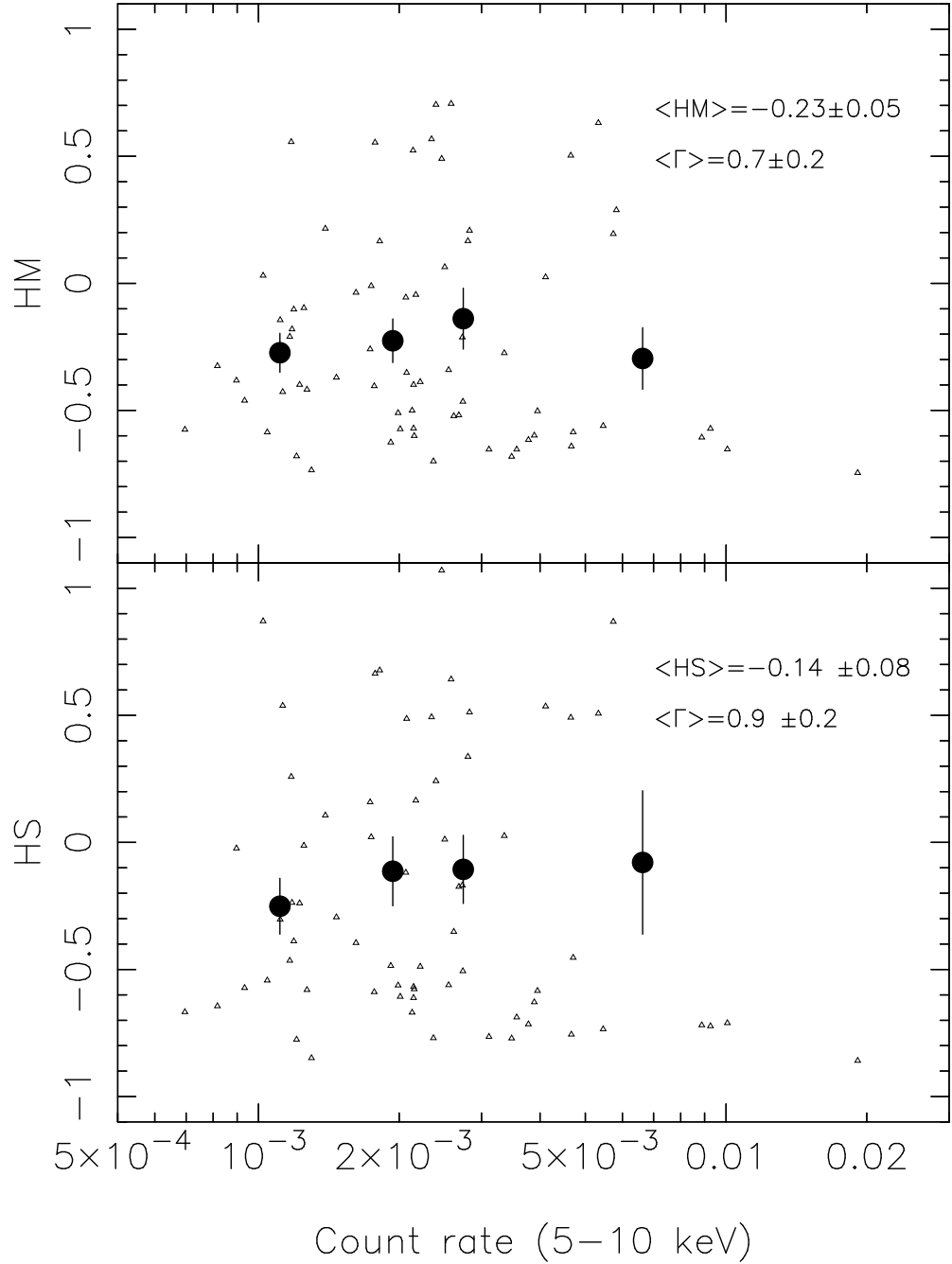}
\caption{
As in Fig.~\ref{fig:hr1}, but with the hardnesses binned
according to 5-10 keV count rate large filled circles) along with the raw
values (small open triangles). The binned values are unweighted, and
the errors determined from the dispersion of the points. There is
no trend for the hardness to change with flux, although the brightest
four objects are clearly and significantly softer than the mean
(Fig.~\ref{fig:hr1}). The mean hardnesses both correspond to a very flat
spectrum, with $\Gamma\sim 0.8$, considerably harder than the spectra
of objects found in 2-10 keV surveys (see text).
\label{fig:hr2}}
\end{figure}

The SHEEP sources are selected in the 5-10 keV (``hard'' = H) band,
but we have also extracted the background subtracted count rates in
the 2-5 keV (``medium''= M) band and the 0.7-2 keV (``soft''=S)
band. These broad-band fluxes allow us to characterize the spectra of
the sources in a crude manner.  In Figs.~\ref{fig:hr1} and
\ref{fig:hr2} we plot the hard/medium ratio $HM=(H-M)/(H+M)$ and the
hard/soft ratio $HS=(H-S)/(H+S)$ versus the hard band count rate.
Note that in determining these ratios we have not corrected the counts
for vignetting and the variable PSF. The energy dependence of these
quantities is very weak (Serlemitsos et al. 1995) and correcting for
these effects introduces an uncertainty larger than the correction. To
the extent that these corrections change the hardness values, the
effect is to make the corrected values slightly larger (i.e. harder).

Several things are noteworthy about Fig.~\ref{fig:hr1}. First, the
hardness values can be used to determine the mean or typical effective
spectral index for the objects. This is necessary for the conversion
of count rates to flux, and also for comparing the typical spectrum of
our objects to the X-ray background and those detected in other
surveys. Determining the mean hardness of the sources is not
straightforward, however, and the inferred spectrum depends on the
method adopted. There are at least three ways of determining the mean
HM and HS values: the unweighted average, the average weighted by the
error bars and stacking.  The last involves adding together all the
individual H and M values (for example) and then computing HM from
these summed fluxes.  We prefer the first method (unweighted average)
for determining the mean source spectrum as the last two can be
strongly skewed by a small number of very bright sources. For example,
the brightest source in our sample has a total number of M counts
comparable to the sum of the entire remainder of the sample. The
weighted and particularly the stacked HM value would therefore have
little meaning, as it is largely representative of the spectrum of
this single source, rather than the whole sample.  The average
hardness ratios of the full sample using all three methods are shown
in Table~\ref{tab:hr}. We have also calculated the average hardness
for subsamples excluding the brightest source, and also the four
brightest sources. Only the unweighted average gives consistent
answers, showing how strongly the other methods can be biased by a
very few bright sources.

Using the unweighted mean has the additional benefit of allowing us
to estimate the error on the mean using the dispersion of the points
which, at least in part, allows for some intrinsic dispersion in
hardness values (see also Maccacaro et al. 1988). The unweighted mean
HM value for our sample corresponds to a spectral index $\Gamma=0.7\pm
0.2$ assuming no absorption. The HS hardness gives a very similar
spectrum, corresponding to $\Gamma=0.9 \pm 0.2$. In practice the
spectra may well be absorbed rather than showing this very flat photon
index, but we note that the mean spectrum of our sources is
significantly flatter than the integrated spectrum of the X-ray
background in the 2-10 keV band, which has effective $\Gamma=1.4$
(e.g. Gendreau et al. 1995). 

In Fig.~\ref{fig:hr1}, we have differentiated between
sources which were detected by ROSAT either in the pointed phase or
RASS, those which were observed in the pointed phase but not detected,
and those which were not observed, except in the RASS. It is clear
that the ROSAT detected sources, as might be expected, exhibit
significantly softer spectra than both the ROSAT non-detections, and
the sample as a whole. They are nonetheless quite hard, with mean HM
equivalent to $\Gamma=1.4$ (Table~\ref{tab:hr}). The ROSAT
non-detections are extremely hard with equivalent $\Gamma \sim 0$
based on the HM color (Fig.~\ref{fig:hr1} and Table~\ref{tab:hr}). Many
of our sources therefore clearly have spectra harder than that of the
XRB. Fig.~\ref{fig:hr2} shows the plot of the binned hardness ratio. This
shows no tendency to harden with at faint fluxes, contrary to the
findings of 2-10 keV surveys (Ueda et al. 1999b; Della Ceca et al.
1999; Giommi et al. 2000; Giacconi et al. 2001). It is clear from
Fig.~\ref{fig:hr1}, however, that the very brightest sources are
anomalously soft. For example, the brightest four objects in the
sample have HM hardness corresponding to $\Gamma=2.1$, much softer
than the mean, but fairly typical for unobscured AGN (e.g.  Nandra \&
Pounds 1994). A total of 21 of the remaining SHEEP objects have a
spectrum harder than this value at the 2$\sigma$ level. 

There are disadvantages of using the straight average of the HM or HS
values to determine the typical spectrum. One is that we give the same
weight to data points that are very poorly determined as those that
are very well determined. This should not matter if a sufficiently
large number of points are averaged, but we note that a relatively
large a fraction of SHEEP sources have hardnesses that are essentially
undefined (i.e. the error on the HM or HS value spans the entire range
of -1 to +1). To test whether these have a biasing effect, we have
excluded from the averaging all such sources. There are 16 in total
(Table~\ref{tab:sample}).  Excluding these does indeed result in a
softer mean spectrum, as when the remaining 53 sources are averaged we
find $\Gamma=1.3 \pm 0.2$ from HM and a very similar value from HS
(Table~\ref{tab:hr}).  This does not necessarily imply that the mean
values from the entire sample are incorrect - it may simply be that
the hardest sources have poorly defined hardness ratios. This is
indeed expected, as very hard sources would be expected to have small
and therefore uncertain M and/or S count rates. A related and
potentially more serious problem is that of the Eddington bias.  As we
have performed the selection in the ``H'' band, sources around the
flux threshold whose H counts randomly fluctuate in the positive
direction will be deemed detections, while negative fluctuations will
not. The M and S counts suffer no such (statistical) bias and
therefore the weakest SHEEP sources should have HM and HS that are
higher than the true values. We have tested this by considering only
the SHEEP which have been detected at $>6\sigma$. The average HM and
HS values of these most significant 34 sources again correspond to a
softer spectrum of ($\Gamma \sim 1.3$).  While this result may on the
face of it seem to contradict the conclusion that the populations do
not harden to faint fluxes, it does not. This is due to the large
range of exposure times and off-axis angles for the SHEEP objects,
such that flux and significance are not equivalent.

To attempt to mitigate both the effects described above, and to
facilitate comparison with 2-10 keV surveys (see discussion) we have
also calculate the hardness ratio denoted as ``HR1'' by, e.g., Ueda et
al. (2001). This is defined in our notation as (H+M-S)/(H+M+S),
i.e. the 2-10 keV vs 0.7-2 keV hardness.  This should be much less
affected by the Eddington bias as the effective area of the GIS is
larger in the 2-5 keV band, and therefore the 2-10 keV counts will not
typically be dominated by the 5-10 keV counts unless the spectrum is
genuinely hard. We find HR1=$0.28\pm 0.05$ ($\Gamma=1.11\pm 0.11$) for
the entire sample of 69 SHEEP.  This does indeed imply a slightly
softer spectrum than that derived from the HM or HS values (although
it is consistent with the latter). Again considering only the 34 most
significant sources in the sample we find $\Gamma=1.3\pm 0.1$ based on
HR1, consistent at $90$ per cent confidence with the mean for the
whole sample and therefore demonstrating the relative lack of bias in
the HR1 ratio. This value is also completely consistent with those
derived from the HM and HS hardnesses for the $>6\sigma$ subsample.

\begin{figure}[t]
\epsscale{1.0}
\plotone{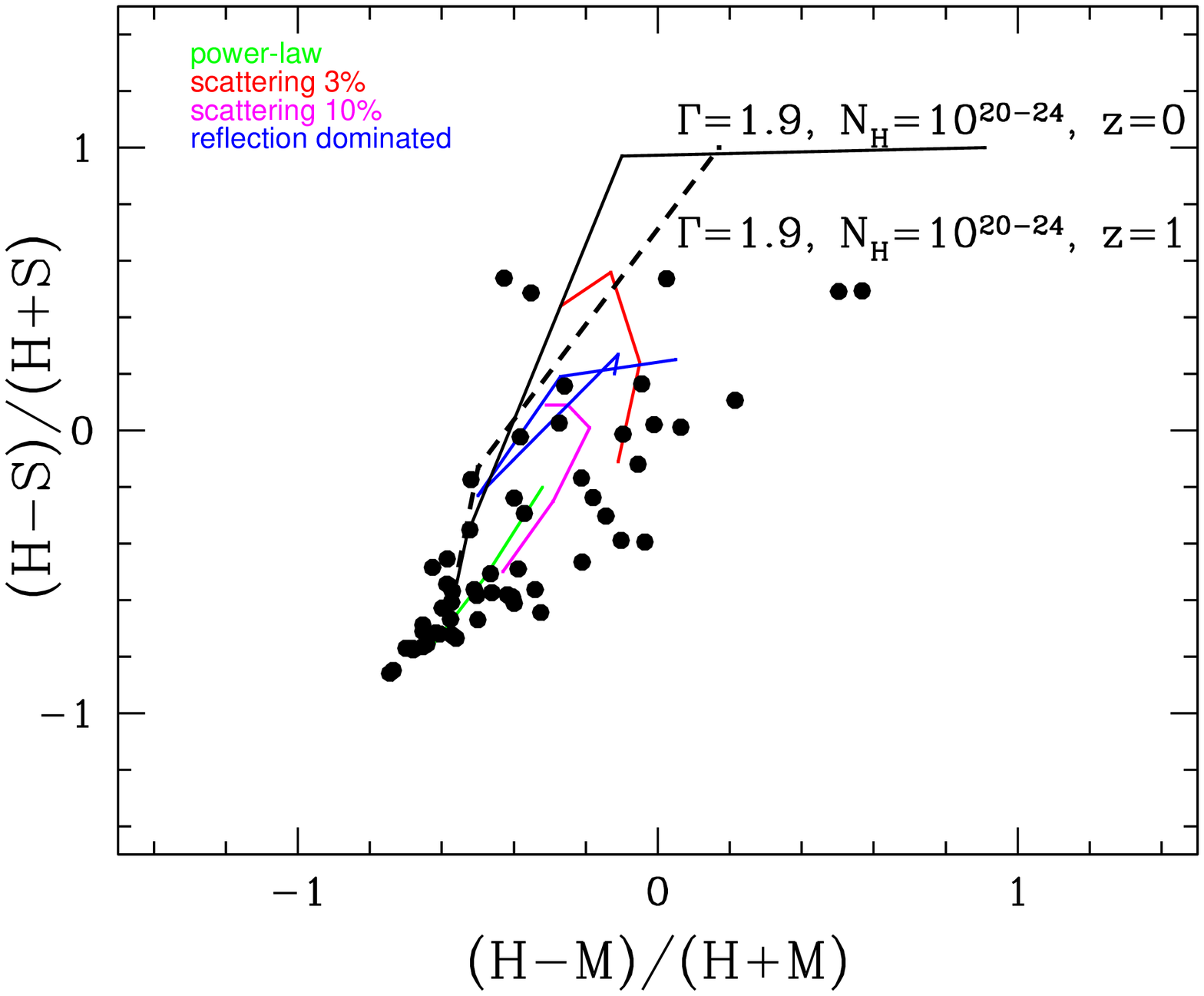}
\caption{The X-ray color-color diagram.  The S, M and H bands
correspond to 0.7-2, 2-5 and 5-10 keV respectively.  Sources with no
detection in the 0.7-2 or 2-5 keV bands have been omitted for
simplicity. The lines represent comparison spectra.  The solid black
line denotes the spectrum of an object at redshift z=0, a photon index
of $\Gamma=1.9$ and a column density of $N_H=10^{20-24}$ $\rm cm^{-2}$
(left to right).  The dashed line denotes the spectrum of an object at
redshift z=1, a photon index of $\Gamma=1.9$ and a column density of
$N_H=10^{20-24}$ $\rm cm^{-2}$ (left to right). The colors are
generally no in good agreement with these simple, absorbed spectra.
The green line represents a power-law spectrum with $\Gamma=1.0-1.9$
(right to left).  The magenta line represents a partial coverer or
scatterer model ($\Gamma=1.9$, column density $10^{24} \rm cm^{-2}$
scattering fraction $f=0.1$) from redshift z=0 to redshift z=2 (top to
bottom). Another such model is shown in red, but this time
with a covering fraction of $f=0.03$ Finally, the blue line represents
the shift of a reflection dominated spectrum with an Fe line (see text
for details) as a function of redshift (z=0 to z=2); the rightmost
point corresponds to z=0. These ``composite'' models are able to
reproduce the colors of a large fraction of the SHEEP sources. 
\label{fig:color}}
\end{figure}

In order to investigate the spectral properties of the individual
sources in more detail, and in particular the role of absorption, we
plot the X-ray color-color diagram in Fig.~\ref{fig:color}, in which
we compare the HM and HS values with those predicted by various
spectral models.  It is evident from Fig.~\ref{fig:color} that there
is a wide range of spectral properties in our sample.  Absorbed model
spectra consisting of a power-law of $\Gamma=1.9$ (the typical AGN
spectrum) and a column $N_H=10^{20-24}$ and z=0 are shown by the solid
black line; at redshift z=1 (dashed line) the spectra are softer as
the K-correction moves the absorption to energies outside the ASCA
band. It is clear that, aside from the softest objects, which have
unabsorbed power law spectrum and cluster at the bottom left of the
diagram, colors of the SHEEP objects are not well represented by these
simple, absorbed spectra. In particular a large fraction show a much
harder HM color, for a given HS, than that for an absorbed power
law. The green line represents a power-law spectrum with no
absorption; the softest end of the line corresponds to $\Gamma=1.9$
while the hardest point is $\Gamma=1.0$. This is more consistent with
the colors but many objects require extremely flat slopes
($\Gamma<1$). The physical origin of such flat power law slopes is
unclear.

It is much more likely that the spectra of the SHEEP are complex, with
both an absorbed hard component and softer emission.  This
``composite'' model is typical of intermediate Seyfert galaxies
(e.g. Seyfert 1.8 -1.9) in the local Universe (e.g. Turner et
al. 1997).  The red and magenta lines corresponds to scattering or
partial covering models in which we assume that the hard X-ray
emission is covered by an obscuring screen of $N_H=10^{24}$ cm$^{-2}$.
Some fraction $f$ of the X-ray emission is either scattered into the
line of sight or represents a ``leaky'' direct component. We evolve
our models from redshift z=0 to z=2, and test two different covering
fractions f=0.03 and f=0.1.  We see that this model provides a good
description for a large number of our sources and in particular is
able to explain the sources which exhibit a very hard HM value, but
relatively soft HS. Della Ceca et al. (1999) have reached similar
conclusions regarding the composite nature of the spectra studying the
spectral properties of a sample of ASCA sources detected in the softer
2-10 keV band. The BeppoSAX 2-10 keV and HELLAS surveys concur (Giommi
et al. 2000; Vignali et al. 2001). The final model (blue line) shown
in Fig~\ref{fig:color} is a reflection dominated spectrum
(e.g. Reynolds et al. 1994; Matt et al. 1996). The intrinsic
illuminating spectrum has $\Gamma=1.9$ but the Compton reflection
component has been enhanced by a factor of 100 compared to a slab
subtending $2\pi$ solid angle at the X-ray source. An iron K$\alpha$
line at 6.4 keV of 1 keV equivalent width has also been
included. Again this model is evolved from z=0 to z=2 and like the composite
spectra, the reflection dominated model predicts harder HM values for
a given HS.  Indeed, the reflection dominated spectrum may be
considered to be the extreme column density end of the ``composite'
model. As is evident from Fig.~\ref{fig:color}, the predicted colors
depend in detail on the column density, scattering (or covering)
fraction and redshift. We can get only loose constraints on these from
our color-color plot, but note that very high column densities ($\gg
10^{23}$~cm$^{-2}$) are required shift the HM colors significantly,
particularly at high redshift.

\begin{figure}[t]
\plotone{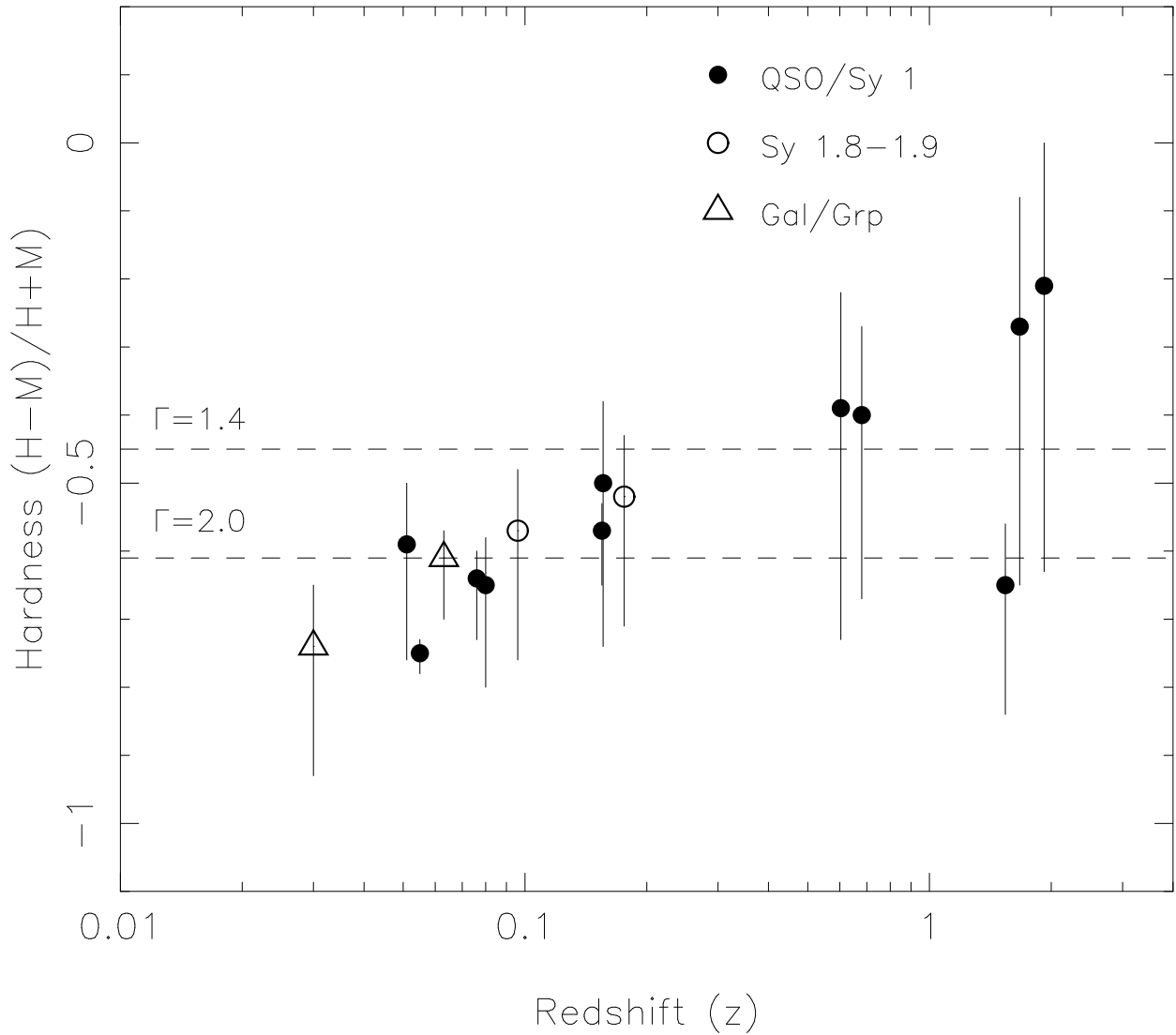}
\caption{HM hardness ratio versus redshift. The horizontal dashed
lines show the HM values corresponding to two relevant values of the
X-ray spectral index. A tentative correlation between hardness and
redshift exists in the data, but with such a small number of objects
we do not consider this to be a robust result. 
\label{fig:hr_z}}
\end{figure}

Fig.~\ref{fig:hr_z} shows the HM hardness ratio versus redshift for
the objects with catalogue identifications (Table~\ref{tab:catids}).
A tentative correlation can be seen between the source hardness and
redshift, with both linear and Spearman rank correlations being
significant at $>99$ per cent confidence.  Clearly such a result would
be of great interest, and similar results have been obtained in HELLAS
(Comastri et al. 2001). We caution, however, that the very small
number of objects in this plot means the true significance of our
result is in some doubt. We therefore defer detailed discussion of the
significance of such a correlation until redshifts are available for a
larger number of the SHEEP sources, which will confirm or deny the
result.

\subsection{Spectral fitting}

\begin{figure}
\plotone{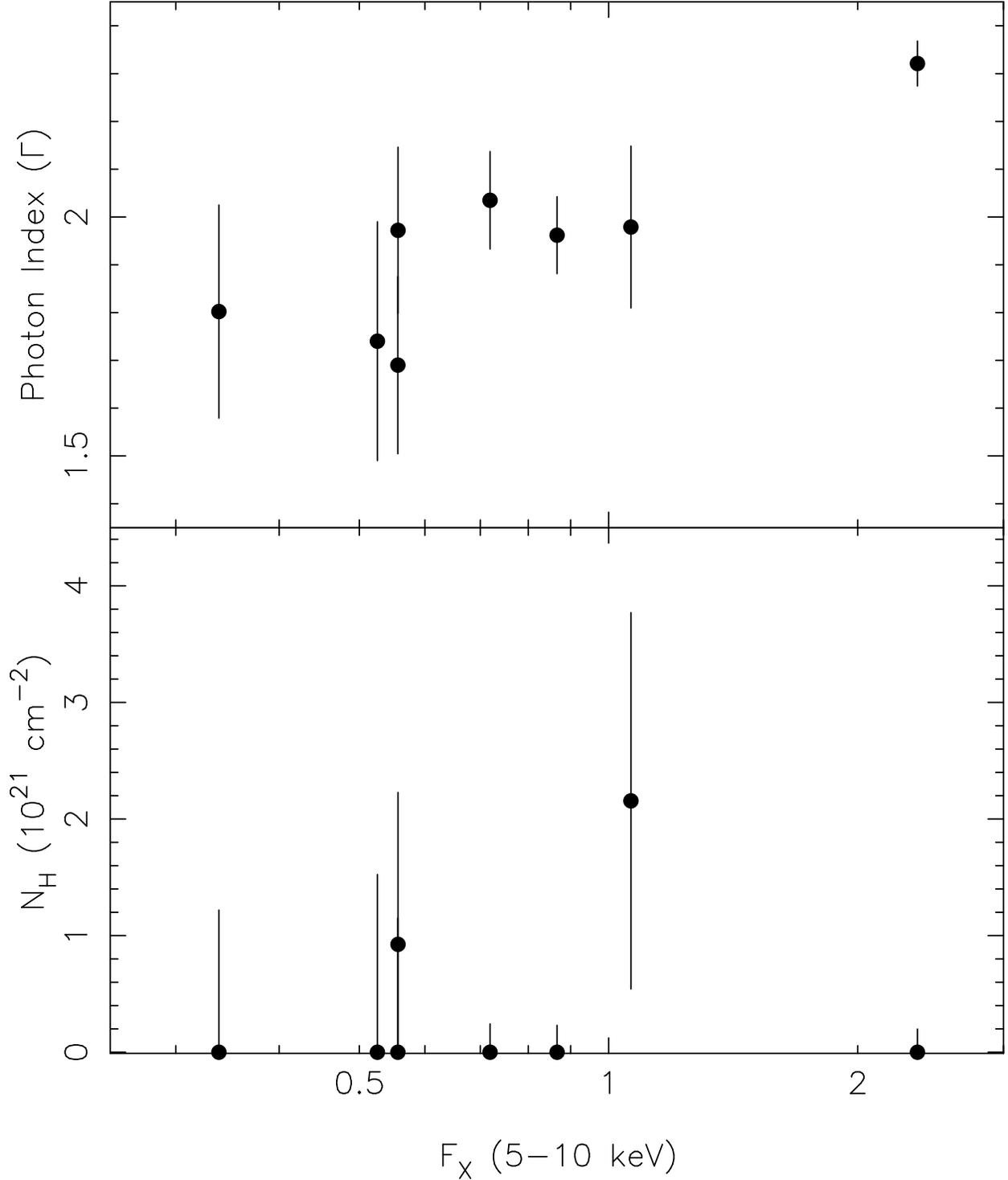}
\caption{Spectral parameters for the 8 SHEEP sources with the
highest signal-to-noise ratio in the 5-10 keV band (S/N$>10$).
The photon index ($\Gamma$; upper panel) and absorbing column density 
(\nh; lower panel) are plotted against the 5-10 keV flux in units
of $10^{-12}$~erg cm$^{-2}$ s$^{-1}$. All
parameters were derived from a spectral fit to the 1-10 keV GIS
data with an absorbed power law model, assuming a minimum of 
Galactic \nh. Only one spectrum shows 
evidence for significant absorption. This may only be revealed in the
spectra of weaker objects, which requires data with higher signal-to-noise
ratio. 
\label{fig:spec}}
\end{figure}

A few of the SHEEP sources are bright enough to allow direct spectral
fitting. Here we consider only the 8 sources in Table~\ref{tab:sample}
which have gaussian S/N in the 5-10 keV band $>10 \sigma$. We
extracted the GIS2 and GIS3 spectra for these sources in the full
band, and grouped them such that they had a minimum of 20 counts in
each bin.  Background was taken from adjacent source-free regions of
the detector.  We then fitted the spectra in the 1-10 keV band with
model of a power law with a free, neutral absorber in the
line-of-sight, in addition to fixed Galactic $N_{\rm H}$. The results
are shown in Table~\ref{tab:fits}.

The spectra are all well fit with this absorbed power law model with
no evidence for any deviation from it. The mean spectral index is
$\Gamma=1.94$ (unweighted) typical of the 2-10 keV spectra of bright,
hard X-ray selected AGN (Nandra \& Pounds 1994). This value is also
similar to that of the soft X-ray background, and soft X-ray selected
QSOs (e.g. Georgantopoulos et al. 1996; Blair et al. 2000). Although
we find no general trend of hardening with decreasing flux in the
sample, the spectral fitting does confirm the fact that the sources
with the highest signal-to-noise ratio in our sample are softer than
the whole.  Only one of the sources with S/N$>10$ shows evidence for
significant absorption in the line-of-sight, AX J2020.3-2226. Even
this is relatively modest, at $2 \times 10^{21}$~\pcmsq, with the
upper limits for the other sources typically being less than this
value.  The softness of these very bright sources may explain why we
typically find slightly softer spectra on average from the hardness
ratio analysis when we split the sample according to signal-to-noise
ratio.  Although, as discussed above, there is not a one-to-one
correspondence between flux and significance, these 8 sources with the
with the highest signal-to-noise ratio are also the brightest ones in
the SHEEP sample.  While we have not found any general trend for
hardening towards faint fluxes, it does therefore appear that the
brightest few sources in the sample are softer than the average.  Firm
conclusions cannot be drawn about the origin of the XRB ``spectral
paradox'' without good quality data on the weaker SHEEP sources, that
are more typical of the sample and present colors consistent with the
XRB spectrum.

\section{DISCUSSION}

We have performed an X-ray survey with the ASCA GIS in the 5-10 keV
band in a $\sim 40$~deg$^{2}$ area to a flux level of $\sim
10^{-13}$~erg cm$^{-2}$ s$^{-1}$. 69 sources were detected, with a
logN-logS distribution consistent with HELLAS, and 2-10 keV BeppoSAX
and \asca\ surveys. We resolve $\sim 15$\% per cent of the hard X-ray
background.  Of the 69 sources, 35 have ROSAT counterparts and 19 of
these have optical counterparts in catalogs.  The classifications show
that 11 of the ROSAT-detected sources are associated with type-1 AGN
(i.e. Seyfert-1 and QSOs). We have shown, however, that the sources
with ROSAT counterparts are preferentially softer than the remainder
of the survey, and are therefore not an unbiased sample. A relatively
large fraction (40\%) of our sources were observed by ROSAT in the
pointed phase, but not detected, and must therefore have extremely
hard spectra.

The mean spectrum of the entire sample, as determined by the mean
hardness ratio, depends on the method adopted and can be affected by
statistical bias. However, we find that it is at least as hard as the
X-ray background, and what should be the least biased estimate gives
an equivalent $\Gamma=1.1\pm 0.1$.  Unlike previous 2-10 keV surveys
(e.g. Ueda et al. 1999b; Della Ceca et al. 1999; Giommi et al. 2000;
Giacconi et al. 2001), we find no systematic hardening of the spectra
to faint fluxes, although direct spectral fitting shows that the
brightest few objects in our sample are anomalously soft.  The spectra
of the bulk of the individual sources are best described by a
composite model, in which the power law is heavily absorbed, but some
fraction of it either leaks through the absorber, or is scattered back
into the line of sight. Some of the sources present very hard X-ray
colors in the 2-10 keV band, much harder than the spectrum of the
XRB. These may be very heavily absorbed, Compton thick sources such as
NGC 6552 (Reynolds et al. 1994) and the Circinus galaxy (Matt et
al. 1996).  This type of object could be quite common, and be a major
contributer to the peak of the XRB spectrum at $\sim 30$~keV.

\subsection{Comparison with HELLAS and 2-10 keV surveys}

A crucial question for our survey is that of whether it selects
different objects than surveys in the 2-10 keV band. As the 2-10 keV
observations are generally more sensitive, there would be no point in
performing harder surveys such as ours if this were indeed the case.
This issue was not addressed in the analysis of the HELLAS data.  A
related question is whether the 5-10 keV survey simply picks out hard
(e.g. flat or absorbed) objects and misses soft ones, or whether 5-10
keV detection selects object in a less biased way. In the standard AGN
synthesis picture where the objects are harder due to absorption we
expect the latter to be the case. At the same equivalent flux limit,
the 5-10 keV survey should pick up all of the unabsorbed objects, but
it should also find absorbed objects that are missed in the softer
surveys. This holds in part because the intrinsic spectral index of
AGN is approximately $\Gamma=2.0$, and therefore equal intrinsic flux
is emitted per unit energy. If the only modifier is absorption, then
at the same flux limit the 5-10 keV survey should pick up all objects
in the 2-10 keV surveys, and in addition all objects missed by the
2-10 keV surveys because their flux is depressed by absorption in the
2-5 keV band.

These issues are partially resolved by our analysis, although our data
are somewhat contradictory. The fact that we (and HELLAS) find a very
similar logN-logS function to the 2-10 keV surveys suggests that we
are sampling the same populations. Two effects indicate that this is
not the case, however. First, the hardness ratio analysis clearly
indicates that the SHEEP sources have significantly harder spectra than
those obtained in 2-10 keV surveys. As we have discussed above, the
mean hardness ratios is rather difficult to calculate in a robust
manner, and is subject to statistical bias, but we can compare our
preferred value of $\Gamma=1.1\pm 0.1$ with that from 2-10 keV
surveys.  Both Della-Ceca et al. (1999) and Ueda et al. (1999) have
presented mean values for 2-10 keV index derived from direct spectral
fitting of stacked spectra. The former find $\Gamma=1.74\pm 0.07$,
taking the weighted average of their ``bright and ``faint''
subsamples, and the latter $\Gamma=1.49 \pm 0.10$ from a 2-10 keV
sample which was flux-selected to ignore the brightest sources. 
The ``faint'' subsample of Della Ceca et al. (1999) is marginally
consistent with our preferred spectrum, with $\Gamma=1.36 \pm 0.14$

Our analysis has highlighted the difficulty in determining the average
spectral properties of sources detected in different ways in different
surveys. To provide what is perhaps the fairest possible comparison,
we have computed the HR1 hardness ratio for a subsample of the ASCA
sources of Ueda et al. (2001). We considered only serendipitous
sources (i.e. not the targets) and truncated their sample at a
detection level of 4.5$\sigma$ (i.e. our detection threshold) in the
2-10 keV band, which resulted in a total of 601 sources. The mean,
unweighted HR1 value of this sample is HR1=$-0.02 \pm 0.01$,
corresponding to $\Gamma=1.75 \pm 0.02$.  The corresponding HR1 for
our sample is stated in Table~\ref{tab:hr} and corresponds to
$\Gamma=1.11\pm 0.11$.  These surveys were performed with the same
instrument, and the hardness ratios were calculated in the same band
and by the same method. The only substantive difference should
therefore be the selection band (2-10 keV in the Ueda et al. subsample
vs. 5-10 keV for SHEEP). We further note that the Eddington bias
should artificially harden the value from the 2-10 keV survey (because
the sources were selected in that band) but not the mean SHEEP
spectrum. Thus we can firmly conclude that, based on the spectral
form, our 5-10 keV survey selects a different and much harder
population than 2-10 keV surveys.

Although the logN-logS functions are similar in the 2-10 keV and 5-10
keV bands, a few more objects can be accomodated in the 5-10 keV
counts, particularly given the uncertainty in spectral shape and
therefore the conversion of counts to flux. To put this on a more
quantitative footing, we have taken the maximum difference in the
logN-logS normalization comparing our and Cagnoni et al.'s 2-10 keV
survey of 15 per cent. Could this additional 15 per cent of objects
harden our average spectrum sufficiently to cause the difference
between our survey and the 2-10 keV samples?  We have tested this by
excluding the hardest 9 sources (i.e. 15\%) based on their HR1 value)
from the SHEEP sample and recomputing the hardness ratio. We find a
mean hardness for these 60 sources corresponding to $\Gamma=1.33\pm
0.07$, still considerably flatter than the 2-10 keV value. Similar
results are found when excluding objects based on their HM and HS
hardnesses.  We can therefore further conclude that the 5-10 keV
survey does not simply pick up a few {\it additional} hard objects
compared to the 2-10 keV surveys, but rather samples a different
population.

Additional supporting evidence for this conclusion comes from the fact
that we find no trend for the source population to harden at faint
fluxes, which is found in the 2-10 keV surveys.  HELLAS similarly
fails to find such a correlation (Fiore et al. 2001), and we also note
that the Chandra survey of Moretti et al. (2002) reveals no
correlation between hardness and 2-10 keV flux, although such a
correlation is observed with the flux in the softer 0.5-2 keV band
(Giacconi et al. 2001; Moretti et al. 2002). While there is some
evidence that the brightest objects in our survey are softer than the
mean, it appears that the 5-10 keV selection methods digs into the
faint, hard populations which make up the X-ray background much more
quickly than the 2-10 keV surveys. Thus, despite the good agreement
between the number counts, we cannot be sampling the same populations
as the 2-10 keV surveys.  Presumably this is also true of HELLAS,
although no mean spectrum has been given for these sources.  We are
then led to the conclusion that the agreement between the 2-10 keV and
5-10 keV number counts is largely coincidental, with our 5-10 keV
survey picking up many additional hard objects, but almost exactly
compensating for this in terms of numbers by losing softer ones.

The fact that the number counts from the 5-10 and 2-10 keV surveys
agree so well, but that the populations are clearly different
spectrally is troubling for the population synthesis models
(e.g. Madau, Ghisselini \& Fabian 1994; Comastri et al. 1995, 2001;
Gilli et al. 1999, 2001). The problem is that these models generally
assume an intrinsic spectrum for AGN of the form $\Gamma=1.9$, typical
of local AGN, or even flatter. The intrinsic flux of such a power law
per unit energy is larger in the 5-10 keV band than in the 2-5 keV
band or for that matter the 2-10 keV band.  Furthermore, absorption is
invoked for a very large fraction of these sources which further
depresses the 2-5 and 2-10 flux relative to the 5-10 keV flux. For
example 75~per cent of sources in the model of Gilli et al. (1999)
have $N_{\rm H} > 10^{23}$~cm$^{-2}$. At z=0, this column density
suppresses the 2-10 keV flux by a factor $\sim 2$, while the 5-10 keV
flux only changes by $\sim 15$ per cent. These numbers are almost
identical for a more typical AGN synthesis source with $N_{\rm
H}=10^{24}$~cm$^{-2}$ at z=1.5. Thus, if the populations synthesis
models are correct, and the absorbed populations have the same
luminosity function and evolutionary properties as the unabsorbed
ones, we expect much higher number counts (perhaps by a factor $\sim
2$) in the 5-10 keV logN-logS than for the converted 2-10 keV. Such a
conclusion is grossly incompatible with our data (Fig.~\ref{fig:xnc}),
and those from HELLAS. We note, however, that Comastri et al. (2001)
have claimed consistency of both the 5-10 keV and 2-10 keV number
counts with the synthesis models. This may in part be due to the fact
that the proportion of absorbed sources in the synthesis models
depends on the flux limit. Specifically, a higher proportion of
absorbed objects is expected at fainter fluxes. In this scenario, the
apparent agreement in number counts means that at the flux limits
probed by SHEEP and HELLAS, the number of heavily obscured sources
must be very small. It is very hard to see how this can be the case
given the strong difference in spectra we find between the 5-10 and
2-10 keV populations.

An alternative possibility is that there is a large population of very
soft sources (unobscured and with $\Gamma > 2.0$ ) which are missed in
the 5-10 keV surveys and picked up at 2-10 keV.  This is not
postulated typically in the synthesis models, where the obscured
populations dominate and where ``soft excesses'', if present, never
affect the spectra above 2 keV.  We await further detailed modelling
of the number counts to address these issues, but note that there have
been some tentative suggestions that much of the X-ray background may
be produced at low redshift ($z<1$; Tozzi et al. 2001; Rosati et
al. 2002), in agreement with our finding that the luminosity function
and/or evolution of the absorbed populations is likely to differ from
that of standard QSOs.

\subsection{The nature of the X-ray background sources}

Chandra deep surveys have resolved most of the X-ray background into
discrete sources (Mushotzky et al. 2000; Giacconi et al. 2001; Brandt
et al. 2001b). The astrophysical nature of these sources remains
mysterious, however. Many of them are extremely faint in the optical
(Mushotzky et al. 2000; Barger et al. 2001; Alexander et al. 2001)
making spectroscopic identification impossible. What is required to
make further progress, then, is to find bright, nearby examples of
these objects that can be detected and studied in more detail. To do
this requires surveys with much larger area than those possible
currently with Chandra. The SHEEP survey, like HELLAS, with its large
area and hard X-ray selection criterion, provides such a
sample. Indeed it is very clear from the failure of ROSAT to detect a
large fraction of our sources (which are nonetheless very bright in
the hard X-ray band), that there are a large number of hard and
probably obscured sources. While these are likely AGN, how their
astrophysics is related to the more familiar classes of Seyfert 1s,
Seyfert 2s and QSOs in the local and soft X-ray universe remains an
open question. Optical identification and detailed X-ray spectroscopy
of our sample can resolve this issue.

One key question is whether our sources present hard spectra due to a
large amount of intrinsic absorption or whether they are intrinsically
hard. The population synthesis models predict the former, but it is
possible that the sources which make up the hard X-ray background have
flat spectra for other reasons, such as the radiation mechanism. For
example, photon-starved Comptonization or bremsstrahlung emission from
an ADAF would produce a spectrum similar to that of the X-ray
background.  A pure reflection spectrum, e.g. in the case of a
Compton-thick Seyfert galaxy could produce an even flatter spectrum
(Reynolds et al. 1994; Matt et al. 1996, 2000). We find no clear
answer to this in our hardness ratio analysis, but the most likely
situation is that the X-ray spectra are composite, with an absorbed,
hard power law and soft emission that may be either scattered nuclear
light, or a separate thermal component (see also Della Ceca et
al. 1999; Giommi et al. 2000; Vignali et al. 2001). Followup
observations of the HELLAS and SHEEP sources with XMM will determine
this unambiguously. 

One early indication from HELLAS was that there may be a population of
``red quasars'' (Webster et al. 1995), with hard and possibly absorbed
X-ray spectra (Fiore et al. 1999; Vignali et al. 2000). Complete
optical followup of the SHEEP sample will confirm this, but here we
highlight another potentially interesting class, of hard QSOs (see
also Comastri et al. 2001). Our cross-correlation with the NED catalog
shows two sources which are classified optically as QSOs, but whose
hardness ratios indicate extremely hard spectra that correspond to
$\Gamma<1.0$ if they are unabsorbed. We do not have complete optical
spectra or spectral energy distributions of these sources, so these
may also be red quasars and absorbed in the optical. The QSO
classification, however, implies that we are seeing the nuclear broad
lines directly. Again these objects may simply have intrinsically hard
spectra, but to flatten a more typical QSO spectrum of $\Gamma \sim
1.9$ to the hard value observed requires a column density $\gg
10^{23}$~\pcmsq. The dust associated with such gas would likely
obliterate the optical/UV emission, including the broad lines, causing
the source to appear as a type II quasar. That the broad lines are in
fact observed implies that the line of sight is not particularly
dusty. One possibility is that the gas-to-dust ratio in these objects
differs substantially from Galactic values (Maiolino et al. 2001).
Another is that we are seeing hot and /or photoionized gas - or ``warm
absorbers'' (Halpern 1984) - at high redshift. Such a gas component is
commonly observed in low redshift Seyferts (e.g. Nandra \& Pounds
1994; Reynolds 1997; George et al. 1998). The low redshift analogue of
these QSOs is the famous Seyfert galaxy NGC 4151, which shows strong
UV emission and broad optical/UV emission lines, but which is heavily
absorbed in the X-ray band. The most obvious explanation for this is
that there is dust-free gas near the nucleus, and this interpretation
is supported by the fact that the X-ray column NGC 4151 is apparently
mildly ionized (Yaqoob, Warwick \& Pounds 1989; Weaver et al. 1994).

\subsection{A complete, hard X-ray selected sample}

Our survey has defined a new, hard X-ray selected sample of AGN. Along
with the HELLAS AGN, these will be the first complete samples since
the HEAO-1 survey (Piccinotti et al. 1982). Many of the objects in the
Piccinotti sample are the most heavily observed AGN and their detailed
study has revealed much of what we know about nuclear activity in
galaxies. Those sources were almost exclusively nearby Seyferts,
however, and our survey has already revealed a large number
higher-redshift and higher-luminosity AGN. Follow-up observations of
these hard X-ray bright objects could revise our opinions of the
central regions of AGN. If, as is suggested by the Chandra data, the
majority of AGN have been missed by optical surveys, much of what we
think we know about their properties could be misleading. With hard
X-ray selection we avoid the biases against obscuration inherent in
most other methods of selection, and if we can follow up these
observations with high quality data in other wavebands, our opinions
about AGN phenomenology could change dramatically.

\subsection{Future work}

Optical followup of our sources is already in progress, with an
imaging program and some spectroscopy, particularly for the northern
sources.  A significant problem, however, is that the GIS positions
are not alone sufficient to identify unambiguously the optical
counterpart. ROSAT PSPC positions are better and HRI positions the
best available, but as we have shown, the ROSAT detected sources
represent a biased subsample, and these sources probably do not
represent the population providing the bulk of the energy density of
the X-ray background at 30 keV.  What is really required is to obtain
Chandra and/or XMM observations of our sources, which will give us the
optical counterparts without ambiguity. Such observations have the
added advantage that they will allow us to determine the X-ray extent
and spectra of the source populations, providing a crucial piece in
the puzzle of how the bulk of the extragalactic background light at
hard X-ray energies is produced.

\acknowledgements

KN and TJT are supported by NASA grants NAG5-7067 and NAG5-7538, which
also support the Tartarus database.  This research has made use of the
NASA/IPAC Extragalactic database, which is operated by the Jet
Propulsion Laboratory, Caltech, under contract with NASA; and of data
obtained through the HEASARC on-line service, provided by NASA/GSFC.
We thank Richard Mushotzky for many enlightening discussions, and the
referee, Andrea Comastri, for many helpful comments which
substantially improved the paper.

\clearpage

\begin{deluxetable}{llrrrrrrcr}
\tabletypesize{\scriptsize}
\tablecolumns{10}
\tablecaption{The SHEEP Sample \label{tab:sample}}
\tablehead{
\colhead{AX} & 
\colhead{RA} &
\colhead{DEC} &
\colhead{S/N} &
\colhead{Raw} & 
\colhead{Rate} & 
\colhead{hm} & 
\colhead{hs} &
\colhead{Seq.} & 
\colhead{Exp} \\
        \colhead{(1)}   &
        \colhead{(2)}   &
        \colhead{(3)}   &
        \colhead{(4)}   &
        \colhead{(5)}   &
        \colhead{(6)}   &
        \colhead{(7)}   &
        \colhead{(8)}   &
        \colhead{(9)}   &
        \colhead{(10)}   
}
\startdata
J$0026.3+1050$    & 00 26 19.0 & $+10$ 50 40 &   6.2 & $   3.0 \pm 0.8$ & 
$    10.4 \pm    2.6$ & $-0.59^{+0.11}_{-0.23}$ & $-0.54^{+0.12}_{-0.25}$ &
74011000 &   86.9 \\
J$0043.7+0054$    & 00 43 47.0 & $+00$ 54 36 &   6.6 & $   3.6 \pm 0.9$ & 
$    11.8 \pm    2.9$ & $-0.18^{+0.18}_{-0.35}$ & $-0.24^{+0.16}_{-0.32}$ &
75020000 &   72.5 \\
J$0044.0+0102$    & 00 44 03.7 & $+01$ 02 27 &   7.4 & $   3.8 \pm 0.9$ & 
$    26.2 \pm    6.1$ & $-0.52^{+0.10}_{-0.21}$ & $-0.35^{+0.14}_{-0.28}$ &
75020000 &   72.5 \\
J$0058.7+3019$    & 00 58 43.0 & $+30$ 19 45 &   4.9 & $   2.4 \pm 0.7$ & 
$    21.4 \pm    6.6$ & $ 0.52^{+0.48}_{-1.52}$ & $ 1.00^{+-0.00}_{-2.00}$ &
74000000 &   79.1 \\
J$0133.7-4303$    & 01 33 47.1 & $-43$ 03 23 &   4.8 & $   2.2 \pm 0.6$ & 
$     7.0 \pm    2.1$ & $-0.58^{+0.15}_{-0.30}$ & $-0.67^{+0.11}_{-0.24}$ &
75074000 &  102.2 \\
J$0140.1+0628$    & 01 40 08.1 & $+06$ 28 06 &   8.3 & $   3.2 \pm 0.6$ & 
$    11.3 \pm    2.1$ & $-0.43^{+0.11}_{-0.22}$ & $ 0.54^{+0.46}_{-1.27}$ &
75042000 &  145.8 \\
J$0144.9-0345$    & 01 44 54.2 & $-03$ 45 19 &   5.9 & $   3.2 \pm 0.8$ & 
$    21.5 \pm    5.8$ & $-0.60^{+0.11}_{-0.22}$ & $-0.58^{+0.11}_{-0.23}$ &
75090000 &   73.1 \\
J$0146.2-0354$    & 01 46 15.0 & $-03$ 54 53 &   4.8 & $   2.5 \pm 0.8$ & 
$    28.0 \pm    8.8$ & $ 0.17^{+0.83}_{-1.17}$ & $ 0.34^{+0.66}_{-1.34}$ &
75090000 &   73.1 \\
J$0207.7+3511$    & 02 07 44.3 & $+35$ 11 14 &   6.6 & $   3.4 \pm 0.8$ & 
$    23.5 \pm    5.8$ & $ 0.57^{+0.43}_{-1.46}$ & $ 0.49^{+0.51}_{-1.49}$ &
75077000 &   77.1 \\
J$0242.8-2326$    & 02 42 51.8 & $-23$ 26 03 &   6.7 & $   3.6 \pm 0.9$ & 
$    17.7 \pm    4.3$ & $-0.40^{+0.13}_{-0.27}$ & $-0.59^{+0.10}_{-0.20}$ &
73005000 &   76.0 \\
J$0335.2-1505$    & 03 35 16.7 & $-15$ 05 55 &   6.7 & $   3.3 \pm 0.8$ & 
$    19.2 \pm    4.7$ & $-0.63^{+0.09}_{-0.19}$ & $-0.49^{+0.12}_{-0.25}$ &
74009000 &   82.5 \\
J$0335.6-3609$    & 03 35 40.4 & $-36$ 09 18 &   5.1 & $   2.5 \pm 0.8$ & 
$    24.0 \pm    7.3$ & $ 0.70^{+0.30}_{-1.70}$ & $0.24^{+0.76}_{-1.24}$ &
72007010 &   75.3 \\
J$0336.9-3616$    & 03 36 58.3 & $-36$ 16 11 &   8.0 & $   3.9 \pm 0.9$ & 
$    35.7 \pm    8.0$ & $-0.65^{+0.09}_{-0.19}$ & $-0.69^{+0.08}_{-0.17}$ &
72007010 &   75.3 \\
J$0401.4+0038$    & 04 01 25.6 & $+00$ 38 49 &   5.1 & $   3.1 \pm 0.9$ & 
$     9.0 \pm    2.7$ & $-0.38^{+0.19}_{-0.38}$ & $-0.02^{+0.30}_{-0.57}$ &
73016000 &   64.8 \\
J$0440.0-4534$    & 04 40 02.9 & $-45$ 34 59 &   8.5 & $   4.1 \pm 0.8$ & 
$    17.3 \pm    3.5$ & $-0.26^{+0.14}_{-0.28}$ & $ 0.16^{+0.22}_{-0.42}$ &
75050000 &   91.6 \\
J$0443.3-2820$    & 04 43 21.5 & $-28$ 20 53 &  32.0 & $  18.6 \pm 1.8$ & 
$    92.6 \pm    9.1$ & $-0.57^{+0.04}_{-0.08}$ & $-0.72^{+0.03}_{-0.05}$ &
75086000 &   62.3 \\
J$0642.9+6751$    & 06 42 57.8 & $+67$ 51 45 &   7.1 & $   2.8 \pm 0.6$ & 
$    25.5 \pm    5.6$ & $-0.34^{+0.14}_{-0.28}$ & $-0.56^{+0.10}_{-0.20}$ &
74003000 &  120.1 \\
J$0836.2+5538$    & 08 36 13.9 & $+55$ 38 48 &   6.3 & $   3.3 \pm 0.8$ & 
$    13.9 \pm    3.5$ & $ 0.22^{+0.78}_{-1.22}$ & $ 0.11^{+0.28}_{-0.52}$ &
74036000 &   79.9 \\
J$0836.6+5529$    & 08 36 36.2 & $+55$ 29 42 &   4.9 & $   2.3 \pm 0.7$ & 
$    25.0 \pm    7.8$ & $ 0.06^{+0.94}_{-1.06}$ & $ 0.01^{+0.99}_{-1.01}$ &
74036000 &   79.9 \\
J$0843.0+5014$    & 08 43 04.5 & $+50$ 14 29 &   4.8 & $   2.9 \pm 0.9$ & 
$    11.8 \pm    3.7$ & $ 0.56^{+0.44}_{-1.56}$ & $0.26^{+0.74}_{-1.26}$ &
75067000 &   63.2 \\
J$0844.8+5004$    & 08 44 50.1 & $+50$ 04 37 &   5.0 & $   2.7 \pm 0.9$ & 
$    57.4 \pm   18.0$ & $ 0.19^{+0.81}_{-1.19}$ & $ 0.87^{+0.13}_{-1.87}$ &
75067000 &   63.2 \\
J$1035.1+3938$    & 10 35 10.1 & $+39$ 38 09 &   5.0 & $   3.2 \pm 1.0$ & 
$    12.5 \pm    3.8$ & $-0.09^{+0.26}_{-0.52}$ & $-0.01^{+0.28}_{-0.55}$ &
72020000 &   56.0 \\
J$1040.3+2045$    & 10 40 23.4 & $+20$ 45 47 &  12.7 & $   8.0 \pm 1.3$ & 
$    39.5 \pm    6.5$ & $-0.50^{+0.07}_{-0.14}$ & $-0.58^{+0.06}_{-0.12}$ &
73040010 &   59.1 \\
J$1107.0-1150$    & 11 07 00.2 & $-11$ 50 45 &   4.8 & $   2.5 \pm 0.8$ & 
$    24.6 \pm    7.9$ & $ 0.49^{+0.51}_{-1.49}$ & $ 1.00^{+0.00}_{-2.00}$ &
75012000 &   69.9 \\
J$1115.3+4043$    & 11 15 20.3 & $+40$ 43 03 &  18.9 & $  10.0 \pm 1.2$ & 
$    46.7 \pm    5.8$ & $-0.64^{+0.04}_{-0.09}$ & $-0.76^{+0.03}_{-0.06}$ &
74035000 &   78.0 \\
J$1115.4+5308$    & 11 15 24.1 & $+53$ 08 07 &   5.9 & $   3.0 \pm 0.8$ & 
$    53.3 \pm   14.6$ & $ 0.63^{+0.37}_{-1.63}$ & $ 0.51^{+0.49}_{-1.51}$ &
74098000 &   70.7 \\
J$1153.7+4619$    & 11 53 47.6 & $+46$ 19 55 &   6.2 & $   3.1 \pm 0.8$ & 
$    21.5 \pm    5.5$ & $-0.57^{+0.11}_{-0.22}$ & $-0.57^{+0.11}_{-0.22}$ &
75056000 &   81.1 \\
J$1218.6+0546$    & 12 18 40.1 & $+05$ 46 17 &   5.6 & $   2.5 \pm 0.6$ & 
$     9.3 \pm    2.4$ & $-0.46^{+0.15}_{-0.30}$ & $-0.57^{+0.12}_{-0.24}$ &
74085000 &  115.4 \\
J$1218.9+2957$    & 12 18 55.7 & $+29$ 57 26 &   8.2 & $   4.6 \pm 1.0$ & 
$    26.8 \pm    5.7$ & $-0.52^{+0.09}_{-0.19}$ & $-0.17^{+0.15}_{-0.30}$ &
71046000 &   72.0 \\
J$1219.4+0643$    & 12 19 28.7 & $+06$ 43 42 &   9.4 & $   6.4 \pm 1.4$ & 
$    31.1 \pm    6.6$ & $-0.65^{+0.07}_{-0.15}$ & $-0.76^{+0.05}_{-0.11}$ &
74074000 &   45.6 \\
J$1220.2+0641$    & 12 20 16.2 & $+06$ 41 44 &  10.9 & $   7.2 \pm 1.4$ & 
$    54.6 \pm   10.8$ & $-0.56^{+0.09}_{-0.17}$ & $-0.73^{+0.05}_{-0.11}$ &
74074000 &   45.6 \\
J$1228.4+1300$    & 12 28 26.2 & $+13$ 00 22 &   4.9 & $   3.5 \pm 1.1$ & 
$    12.1 \pm    3.9$ & $-0.68^{+0.10}_{-0.22}$ & $-0.78^{+0.08}_{-0.16}$ &
74051000 &   46.0 \\
J$1230.8+1433$    & 12 30 51.9 & $+14$ 33 23 &   5.0 & $   2.5 \pm 0.8$ & 
$    34.8 \pm   10.8$ & $-0.68^{+0.12}_{-0.24}$ & $-0.77^{+0.09}_{-0.18}$ &
75031000 &   69.4 \\
J$1231.5+1422$    & 12 31 33.5 & $+14$ 22 50 &   5.5 & $   3.1 \pm 0.9$ & 
$     8.2 \pm    2.3$ & $-0.33^{+0.19}_{-0.38}$ & $-0.64^{+0.10}_{-0.21}$ &
75031000 &   69.4 \\
J$1241.3+3501$    & 12 41 21.7 & $+35$ 01 01 &   8.6 & $   4.7 \pm 1.0$ & 
$    14.7 \pm    3.0$ & $-0.37^{+0.12}_{-0.24}$ & $-0.29^{+0.13}_{-0.27}$ &
75081000 &   73.3 \\
J$1243.8+1305$    & 12 43 50.0 & $+13$ 05 17 &   5.9 & $   2.8 \pm 0.8$ & 
$    28.3 \pm    7.8$ & $ 0.21^{+0.79}_{-1.21}$ & $0.51^{+0.49}_{-1.86}$ &
75045010 &   75.9 \\
J$1257.6+3525$    & 12 57 40.1 & $+35$ 25 34 &   4.8 & $   2.6 \pm 0.8$ & 
$    16.2 \pm    4.9$ & $-0.03^{+0.32}_{-0.62}$ & $-0.40^{+0.19}_{-0.39}$ &
75078000 &   74.2 \\
J$1258.4+3528$    & 12 58 29.6 & $+35$ 28 16 &   5.8 & $   2.9 \pm 0.8$ & 
$    27.3 \pm    7.6$ & $-0.21^{+0.21}_{-0.42}$ & $-0.17^{+0.23}_{-0.44}$ &
75078000 &   74.2 \\
J$1325.8-3920$    & 13 25 50.0 & $-39$ 20 40 &   4.7 & $   1.6 \pm 0.5$ & 
$    12.7 \pm    3.7$ & $-0.42^{+0.15}_{-0.30}$ & $-0.58^{+0.11}_{-0.23}$ &
75002000 &  152.9 \\
J$1354.0+3346$    & 13 54 01.0 & $+33$ 46 27 &   6.2 & $   3.3 \pm 0.9$ & 
$    46.5 \pm   12.0$ & $ 0.50^{+0.45}_{-1.50}$ & $ 0.49^{+0.51}_{-1.49}$ &
75068000 &   73.3 \\
J$1354.1+3341$    & 13 54 11.6 & $+33$ 41 03 &   6.7 & $   3.4 \pm 0.9$ & 
$    27.4 \pm    6.8$ & $-0.47^{+0.13}_{-0.27}$ & $-0.51^{+0.13}_{-0.26}$ &
75068000 &   73.3 \\
J$1405.4+2223$    & 14 05 26.9 & $+22$ 23 27 &   7.1 & $   3.8 \pm 0.9$ & 
$    21.3 \pm    5.2$ & $-0.50^{+0.12}_{-0.24}$ & $-0.67^{+0.08}_{-0.17}$ &
72021000 &   65.7 \\
J$1406.1+2233$    & 14 06 08.3 & $+22$ 33 02 &   5.3 & $   2.9 \pm 0.9$ & 
$    17.4 \pm    5.2$ & $-0.01^{+0.31}_{-0.60}$ & $ 0.02^{+0.33}_{-0.62}$ &
72021000 &   65.7 \\
J$1406.2+2228$    & 14 06 13.6 & $+22$ 28 22 &   4.8 & $   2.7 \pm 0.8$ & 
$    10.2 \pm    3.2$ & $ 0.03^{+0.97}_{-1.03}$ & $ 0.87^{+0.13}_{-1.87}$ &
72021000 &   65.7 \\
J$1425.2+2303$    & 14 25 13.7 & $+23$ 03 19 &   4.9 & $   3.4 \pm 1.1$ & 
$    18.2 \pm    5.8$ & $ 0.17^{+0.83}_{-1.17}$ & $ 0.68^{+0.32}_{-1.68}$ &
73078000 &   49.5 \\
J$1426.8+2619$    & 14 26 52.1 & $+26$ 19 35 &   6.6 & $   3.0 \pm 0.8$ & 
$    23.7 \pm    5.9$ & $-0.70^{+0.08}_{-0.16}$ & $-0.77^{+0.06}_{-0.12}$ &
74073000 &   83.4 \\
J$1426.9+2334$    & 14 26 54.3 & $+23$ 34 58 &   4.8 & $   2.2 \pm 0.7$ & 
$    41.1 \pm   12.9$ & $ 0.02^{+0.98}_{-1.02}$ & $ 0.54^{+0.46}_{-1.54}$ &
76060000 &   82.8 \\
J$1428.1+2337$    & 14 28 08.2 & $+23$ 37 40 &   5.9 & $   2.7 \pm 0.7$ & 
$    37.8 \pm   10.3$ & $-0.62^{+0.09}_{-0.90}$ & $-0.72^{+0.07}_{-0.15}$ &
76060000 &   82.8 \\
J$1429.7+4240$    & 14 29 45.0 & $+42$ 40 41 &   6.6 & $   4.1 \pm 1.1$ & 
$    33.6 \pm    8.9$ & $-0.27^{+0.19}_{-0.38}$ & $ 0.02^{+0.27}_{-0.52}$ &
71044000 &   51.9 \\
J$1500.1+3325$    & 15 00 09.2 & $+33$ 25 06 &   5.2 & $   2.7 \pm 0.8$ & 
$    58.3 \pm   17.1$ & $ 0.29^{+0.71}_{-1.29}$ & $ 1.00^{+0.00}_{-2.00}$ &
75082000 &   75.0 \\
J$1511.7+0758$    & 15 11 43.6 & $+07$ 58 54 &   5.4 & $   2.8 \pm 0.8$ & 
$    11.9 \pm    3.3$ & $-0.01^{+0.22}_{-0.43}$ & $-0.39^{+0.15}_{-0.30}$ &
71004000 &   85.8 \\
J$1511.7+5702$    & 15 11 47.9 & $+57$ 02 42 &   5.6 & $   2.5 \pm 0.7$ & 
$    17.8 \pm    4.8$ & $ 0.55^{+0.45}_{-1.55}$ & $ 0.66^{+0.34}_{-1.66}$ &
73080000 &   98.5 \\
J$1512.0+5708$    & 15 12 04.4 & $+57$ 08 05 &   5.2 & $   2.2 \pm 0.6$ & 
$    21.7 \pm    6.3$ & $-0.04^{+0.33}_{-0.64}$ & $ 0.17^{+0.83}_{-1.17}$ &
73080000 &   98.5 \\
J$1531.8+2414$    & 15 31 51.1 & $+24$ 14 43 &   8.5 & $   3.5 \pm 0.7$ & 
$    21.5 \pm    4.1$ & $-0.40^{+0.11}_{-0.22}$ & $-0.61^{+0.07}_{-0.15}$ &
75055000 &  130.7 \\
J$1531.9+2420$    & 15 31 56.5 & $+24$ 20 22 &  11.4 & $   4.4 \pm 0.7$ & 
$    38.9 \pm    6.2$ & $-0.60^{+0.07}_{-0.13}$ & $-0.63^{+0.06}_{-0.13}$ &
75055000 &  130.7 \\
J$1532.3+2401$    & 15 32 19.1 & $+24$ 01 13 &   6.3 & $   2.6 \pm 0.6$ & 
$    12.2 \pm    2.9$ & $-0.40^{+0.14}_{-0.29}$ & $-0.24^{+0.17}_{-0.35}$ &
75055000 &  130.7 \\
J$1532.5+2415$    & 15 32 33.1 & $+24$ 15 13 &   8.6 & $   3.4 \pm 0.6$ & 
$    20.7 \pm    4.0$ & $-0.35^{+0.12}_{-0.24}$ & $ 0.49^{+0.51}_{-1.22}$ &
75055000 &  130.7 \\
J$1545.2+4855$    & 15 45 13.6 & $+48$ 55 06 &   4.8 & $   2.2 \pm 0.7$ & 
$    25.8 \pm    8.0$ & $ 0.71^{+0.29}_{-1.71}$ & $ 0.64^{+0.36}_{-1.64}$ &
75059000 &   83.7 \\
J$1617.0+3506$    & 16 17 05.9 & $+35$ 06 38 &   5.8 & $   2.3 \pm 0.6$ & 
$    20.7 \pm    5.4$ & $-0.05^{+0.27}_{-0.52}$ & $-0.12^{+0.25}_{-0.48}$ &
75000000 &  110.3 \\
J$1617.2+3454$    & 16 17 15.5 & $+34$ 54 36 &   5.1 & $   2.2 \pm 0.6$ & 
$    13.0 \pm    3.7$ & $-0.74^{+0.09}_{-0.19}$ & $-0.85^{+0.06}_{-0.11}$ &
75000000 &  110.3 \\
J$1618.1+3459$    & 16 18 10.7 & $+34$ 59 46 &   5.1 & $   2.3 \pm 0.6$ & 
$    11.7 \pm    3.3$ & $-0.21^{+0.24}_{-0.47}$ & $-0.46^{+0.16}_{-0.33}$ &
75000000 &  110.3 \\
J$1728.2+5013$    & 17 28 13.9 & $+50$ 13 28 &  57.6 & $  28.3 \pm 2.0$ & 
$   191.3 \pm   13.2$ & $-0.75^{+0.02}_{-0.03}$ & $-0.86^{+0.01}_{-0.02}$ &
73022000 &   79.5 \\
J$1746.8+6836$    & 17 46 52.8 & $+68$ 36 17 &  23.8 & $  12.7 \pm 1.5$ & 
$    88.7 \pm   10.1$ & $-0.61^{+0.04}_{-0.09}$ & $-0.72^{+0.03}_{-0.06}$ &
74033000 &   69.8 \\
J$1749.8+6823$    & 17 49 50.1 & $+68$ 23 19 &   9.5 & $   5.0 \pm 1.0$ & 
$    47.1 \pm    9.4$ & $-0.59^{+0.09}_{-0.17}$ & $-0.45^{+0.11}_{-0.22}$ &
74033000 &   69.8 \\
J$1804.5+6938$    & 18 04 30.7 & $+69$ 38 01 &   5.0 & $   2.2 \pm 0.7$ & 
$    22.2 \pm    6.5$ & $-0.39^{+0.17}_{-0.34}$ & $-0.49^{+0.14}_{-0.29}$ &
74086000 &   94.8 \\
J$1808.0+6948$    & 18 08 05.9 & $+69$ 48 21 &   7.1 & $   3.2 \pm 0.7$ & 
$    20.1 \pm    4.6$ & $-0.57^{+0.09}_{-0.19}$ & $-0.61^{+0.09}_{-0.17}$ &
74086000 &   94.8 \\
J$1850.6-7838$    & 18 50 38.7 & $-78$ 38 31 &   8.8 & $   3.1 \pm 0.6$ & 
$    19.9 \pm    3.7$ & $-0.51^{+0.09}_{-0.19}$ & $-0.56^{+0.08}_{-0.17}$ &
75008000 &  153.9 \\
J$2002.7-3300$    & 20 02 47.1 & $-33$ 00 15 &   5.2 & $   2.8 \pm 0.8$ & 
$    11.1 \pm    3.2$ & $-0.14^{+0.24}_{-0.47}$ & $-0.30^{+0.19}_{-0.39}$ &
73001000 &   75.2 \\
J$2020.3-2226$    & 20 20 19.6 & $-22$ 26 48 &  20.8 & $   8.1 \pm 0.9$ & 
$   100.6 \pm   11.4$ & $-0.65^{+0.04}_{-0.08}$ & $-0.71^{+0.03}_{-0.07}$ &
73075000 &  118.1 \\
\tablecomments{
Col.(1): SHEEP object
Col.(2): ASCA RA (2000)
Col.(3): ASCA DEC (2000)
Col.(4): Equivalent ``sigma'' (see text)
Col.(5): Total GIS2+GIS3 counts in the 5-10 keV band divided by total
exposure (GIS2+GIS3) in units of $10^{-4}$ ct s$^{-1}$ and $1\sigma$
uncertainty.
Col.(6): Hard-band (5-10 keV) count rate and error corrected to
on-axis value for a single GIS detector in units of $10^{-4}$ ct
s$^{-1}$. These should be multipled by $1.24 \times 10^{-10}$ to
convert to flux in units of erg cm$^{-2}$ s$^{-1}$ for a $\Gamma=1.6$
spectrum.
Col.(7): Hardness, defined as $(H-M/H+M)$ where $M$ is the medium energy (2-5 keV) rate.
Col.(8): Hardness, defined as $(H-S/H+S)$ where $S$ is the soft (0.7-2 keV) rate.
Col.(9): Sequence number in which this source was detected
Col.(10): Total GIS+GIS3 exposure time for sequence (ks)
}

\enddata
\end{deluxetable}

\clearpage

\begin{deluxetable}{llllllrcr}
\tabletypesize{\scriptsize}
\tablecolumns{9}
\tablecaption{ROSAT observations of SHEEP sources \label{tab:rosat}}
\tablehead{
\colhead{AX} & 
\colhead{Seq} &
\colhead{Exp} & 
\colhead{Theta} & 
\colhead{RX} &
\colhead{Offset} &
\colhead{Rate} &
\colhead{$N_{\rm H}$(Gal)} & 
\colhead{Flux} \\
        \colhead{(1)}   &
        \colhead{(2)}   &
        \colhead{(3)}   &
        \colhead{(4)}   &
        \colhead{(5)}   &
        \colhead{(6)}   &
        \colhead{(7)}   &
        \colhead{(8)}   &
        \colhead{(9)}   
}
\startdata
J0026.3+1050 &  ROSHRI      & 30.0 & 9.2 & J$002619.1+105021$  & 0.31 &
   0.0053 $\pm$ 0.0006 & 6.0 &  1.63 \\
J0043.7+0054 &  rp700377n00 & 10.7 & 39.2 & \nodata            & \nodata &
 $<0.01$ & 2.3 &  $<3.1$  \\
J0044.0+0102 &  WGACAT      & 10.7 & 33.9 & J$004401.4+010258$ & 0.77 &
 0.0095 $\pm$ 0.0019 & 2.4 & 1.0  \\
J0058.7+3019 &  rh701308n00 & 14.9 & 13.0 & \nodata            & \nodata &
 $<0.0060$ & 5.9 & $<5.85$ \\
J0133.7-4303 &  ROSHRI      & 2.4 & 8.2 & J$013346.6-430417$   & 0.90 &
        0.0097 $\pm$ 0.0025 & 1.8 &  3.06  \\
J0140.1+0628 &  ROSHRI   & 4.3 & 8.0 & J$014002.5+062732$      & 1.51 &
 0.0035 $\pm$ 0.0013 & 4.1 & 1.14  \\
J0207.7+3511 &  WGACAT      & 14.0 & 16.1 & J$020744.1+351142$  & 0.47 &
 0.0025 $\pm$ 0.0006 & 6.3 &  0.3   \\
J0242.8-2326 &  ROSHRI      & 20.5 & 6.5 & J$024252.3-232633$  & 0.51 &
 0.0142 $\pm$ 0.0009 & 2.0 &  4.40  \\
J0335.6-3609 &  rp700921a01 & 7.7 & 22.6 & \nodata  & \nodata &
  $<0.0125$ & 1.4 &  $<3.5$ \\
J0336.9-3616 &  ROSHRI      & 6.2 & 16.7 & J$033653.9-361617$  & 0.89 &
 0.0099  $\pm$ 0.0023 & 1.4 & 3.1  \\
J0401.4+0038 &  WGACAT      & 7.8 & 26.9 & J$040125.9+003935$ & 0.77 &
 0.0061 $\pm$  0.0013 & 13.0 & 0.73   \\
J0440.0-4534 &  rh702888n00 & 4.4 & 4.6 & \nodata & \nodata &
 $<0.0012$ & 2.0 & $<1.1$   \\
J0443.3-2820 &  RASSBSC     & 0.1  & \nodata  & J$044320.8-282039$         & 0.28 &
  0.15 $\pm$ 0.04  &  2.4   & 15.9   \\
J0642.9+6751 &  WGACAT      & 5.3 & 7.5 & J$064246.2+675222$ & 1.27 &
 $0.0046 \pm 0.001$ & 5.5 & 0.55      \\
J0836.2+5538 & rh703892n00 & 16.7 & 10.0 & \nodata & \nodata &
 $<0.0062$ & 4.1 &  $<5.8$  \\
J0836.6+5529 & rh703892n00 & 16.7 & 13.3 & \nodata & \nodata &
 $<0.0071$ & 4.1 & $<6.8$    \\
J0843.0+5014 &  WGACAT      & 8.7 & 4.0 & J$084315.3+501415$ & 1.74 &
 0.0032 $\pm$ 0.0007 & 3.1 & 0.35  \\
J0844.8+5004 & rp700318n00 & 8.7 & 14.0 & \nodata & \nodata &
 $<0.0094$ & 3.0 & $<3.0$   \\
J1035.1+3938 & rh701982n00 & 1.6 & 4.5 & \nodata & \nodata &
 $<0.0016$ & 1.5 & $<1.50$     \\
J1040.3+2045 & RASSFSC     & 0.36 & \nodata  & J$104027.5+204555$   & 1.02 &
  0.10 $\pm$  0.02 & 2.0  & 10.2    \\
J1115.3+4043 &  ROSHRI      & 28.9 & 10.6 & J$111521.2+404327$ & 0.42 &
 0.0404 $\pm$ 0.0013 & 1.9 & 12.6   \\
J1153.7+4619 &  WGACAT  & 4.4 & 7.8 & J$115345.6+462022$ & 0.56 &
 0.0107$\pm$ 0.0017 &   2.0   & 1.12    \\
J1218.6+0546 & rh701307a01 & 18.5 & 12.4 & \nodata & \nodata &
 $<0.0055$ & 1.5 &  $<5.0$  \\
J1218.9+2957 &  WGACAT          & 3.1  & 11.7  & J$121854.8+295835$ & 1.16 &
 0.0039 $\pm$ 0.0014 & 1.7  & 0.44  \\
J1219.4+0643 &  ROSHRI      & 3.8 & 6.1 & J$121930.8+064339$ & 0.48 &
   0.0580 $\pm$ 0.0040 & 1.6 &  18.2  \\
J1220.2+0641 &  ROSHRI      & 4.0 & 14.4 & J$122018.3+064123$ & 0.59 &
  0.0482   $\pm$ 0.0042 & 1.6 & 15.0    \\
J1228.4+1300 & rh701657n00 & 5.9 & 8.8 & \nodata & \nodata &
 $<0.0011$ & 2.6 &  $<1.06$   \\
J1230.8+1433 &  ROSHRI      & 4.4 & 16.5 & J$123052.4+143305$ & 0.32 &
 0.0145   $\pm$ 0.0029  & 2.5 &  4.8  \\
J1231.5+1422 & rh601003n00 & 4.4 & 4.1 & \nodata & \nodata &
 $<0.0099$ & 2.5 &  $<9.6$    \\
J1241.3+3501 &  ROSHRI      & 46.3 & 6.0 & J$124123.9+350014$ & 0.92 &
  0.0011  $\pm$ 0.0003 & 1.4  &   0.31\\
J1243.8+1305 & rh701007a01 & 6.1 & 17.4 & \nodata & \nodata &
 $<0.0075$ & 2.3 &  $<7.0$   \\
J1257.6+3525 &  WGACAT      & 4.0 & 7.7   & J$125745.4+352542$  & 1.08 &
 0.0035 $\pm$ 0.0012 & 1.2 & 0.32     \\
J1258.4+3528 &  WGACAT      & 17.2 &  9. & J$125829.4+352840$ & 0.40 &
   0.0252 $\pm$ 0.0023   & 1.2 &   2.28 \\
J1354.1+3341 &  RASSFSC & 0.4 & \nodata  &   J$135407.8+334039$ & 0.85 &
 0.0298 $\pm$ 0.0107    &  1.2   & 2.64   \\
J1405.4+2223 &  ROSHRI      & 1.6 & 10.4 & J$140528.2+222321$ & 0.35 &
 0.0172 $\pm$ 0.0037 & 2.1 & 5.4    \\
J1406.1+2233 & rh703968n00 & 1.7 & 11.3 & \nodata & \nodata &
 $<0.00195$ & 2.1 &  $<1.90$               \\
J1406.2+2228 & rh703968n00 & 1.7 & 4.2 &  \nodata & \nodata &
 $<0.00177$ & 2.1 &  $<1.71$           \\
J1425.2+2303 & rh701899n00 & 5.7 & 9.0 &  \nodata & \nodata &
 $<0.0092$ & 2.7 & $<12.4$    \\
J1426.8+2619 &  WGACAT      & 6.7 & 16.3 & J142652.5+261922$$ & 0.24 &
 0.033 $\pm$ 0.0027 & 1.7 &  3.24  \\
J1426.9+2334 &  WGACAT      & 3.0 & 10.0 & J$142656.1+233651$ & 1.93 &
 0.0045 $\pm$ 0.00015 & 2.7 & 0.49   \\
J1428.1+2337 &  WGACAT      & 3.1 & 18.3 & J$142807.4+233725$ & 0.30 &
 0.0560 $\pm$ 0.0048 & 2.8 & 6.11  \\
J1429.7+4240 &  WGACAT      & 9.5 & 13.7 & J$142944.7+424106$ & 0.42 &
 0.0149 $\pm$ 0.0016  & 1.4 &  1.41 \\
J1512.0+5708 & rp600190n00 & 18.1 & 53.6 & \nodata & \nodata &
 $<0.0053$ & 1.5 & $<1.43$     \\
J1532.3+2401 & rp701411n00 & 23.0 & 52.0 & \nodata & \nodata &
 $<0.0061$ & 4.1 & $<2.1$    \\
J1532.5+2415 & rp701411n00 & 23.0 & 49.7 & \nodata & \nodata &
 $<0.0040$ & 4.1 & $<1.40$    \\
J1545.2+4855 & rp700809n00 & 5.5 & 13.0 & \nodata & \nodata &
 $<0.0013$ & 1.6 &   $<0.38$    \\
J1617.0+3506 & rh800164n00 & 33.4 & 7.9 & \nodata & \nodata &
 $<0.0041$ & 1.4 &   $<3.7$   \\
J1617.2+3454 &  ROSHRI      & 33.4 & 7.0 & J$161720.5+345404$ & 1.12 &
 0.0047 $\pm$ 0.0005 & 1.4 & 1.46  \\
J1618.1+3459 & rh800164n00 & 33.4 & 3.8 & \nodata & \nodata &
 $<0.0038$ & 1.4 & $<3.7$     \\
J1728.2+5013 &  ROSHRI      & 1.6 & 1.4 & J$172819.1+501309$ & 0.90 &
  0.6068 $\pm$ 0.0192  & 2.7 & 437.0  \\
J1746.8+6836 &  WGACAT      & 24.7 & 10.6 & J$174658.2+683632$ & 0.56 &
 0.2790 $\pm$ 0.0039 & 4.4 & 32.7   \\
J1749.8+6823 &  WGACAT      & 24.7 & 19.8 & J$174949.0+682321$ & 0.12 &
 0.0094 $\pm$ 0.0008 & 4.5 &  1.1   \\
J1804.5+6938 &  ROSHRI      & 18.2 & 16.9 & J$180434.4+693734$ & 0.55 &
 0.0065 $\pm$ 0.0013  & 4.5 & 2.1   \\
J1808.0+6948 &  ROSHRI      & 18.2 & 8.0 & J$180813.9+694806$ & 0.73 &
 0.0036 $\pm$ 0.0006 & 4.8 &  1.17   \\
J1850.6-7838 &  WGACAT      & 2.2 & 10.6 & J$185028.8-783814$ & 0.56 &
 0.0057 $\pm$ 0.0019 & 9.2 & 0.69  \\
J2020.3-2226 &  ROSHRI      & 6.4 & 6.7 & J$202021.5-222554$ & 1.00 &
 0.0057 $\pm$ 0.0011 & 6.1 & 1.85 \\
\tablecomments{
Col.(1): SHEEP object;
Col.(2): ROSAT sequence or catalog;
Col.(3): ROSAT exposure (ks);
Col.(4): Off-axis angle (arcmin);
Col.(5): ROSAT ID;
Col.(6): Offset from ASCA position (arcmin);
Col.(7): PSPC or HRI count rate in the 0.1-2 keV band (ct s$^{-1}$);
Col.(8): Galactic $N_{\rm H}$ ($10^{21}$~cm$^{-2}$); 
Col(9): ROSAT flux in the 0.1-2.0 keV band in units
of $10^{-13}$~erg cm$^{-2}$ s$^{-1}$;
}
\enddata
\end{deluxetable}

\begin{deluxetable}{lllcclrcccc}
\tabletypesize{\scriptsize}
\tablecolumns{10}
\tablecaption{Catalog and literature IDs \label{tab:catids}}
\tablehead{
\colhead{AX} & 
\colhead{NED} &
\colhead{RA} &
\colhead{DEC} &
\colhead{Class} &
\colhead{$z$} &
\colhead{Offset} & 
\colhead{Rad} &
\colhead{IR} &
\colhead{Ref} 
\\
        \colhead{(1)}   &
        \colhead{(2)}   &
        \colhead{(3)}   &
        \colhead{(4)}   &
        \colhead{(5)}   &
        \colhead{(6)}   &
        \colhead{(7)}   &
        \colhead{(8)}   &
        \colhead{(9)}   &
        \colhead{(10)}   
}
\startdata
J$0242.8-2326$& FHC93 0240-2339A & 02 42 51.9  & $-23$ 26 34 & QSO  & 0.68 & 0.1 & N & N & 2 \\  
J$0336.9-3616$& FCSS J033654.0-361606 & 03 36 54.0  & $-36$ 16 07 & QSO  & 1.54 & 0.2 & Y & N & 3 \\  
J$0443.3-2820$& HE0441-2826     & 04 43 20.7   & -28   20 52 & QSO  & 0.155 & 0.2 & N & N & 4 \\ 
J$0642.9+6751$& NRRFJ064241.3+675257 & 06 42 41.3  & $+67$ 52 57 & G/Cl & \nodata & 0.7 & N & N & 5\\  
J$1115.3+4043$& 2MASXi J1115208+40432 & 11 15 20.8  & $+40$ 43 26 & Sy1  & 0.076 & 0.1 & N & Y & 6 \\ 
J$1218.9+2957$& 1SAX J1218.9+2958 & 12 18 52.5  & +29 59 01 & Sy1.9 & 0.176 & 0.7 & N & N & 7  \\  
J$1219.4+0643$& MS 1217.0+0700 & 12 19 31.0  & $+06$ 43 35 & Sy1  & 0.08 & 0.1 & N & N & 3 \\ 
J$1230.8+1433$& VPC0774 & 12 30 52.7  & $+14$ 33 03 & G & \nodata & 0.1 & N & N & 8 \\ 
J$1258.4+3528$& FBQS J125829.6+352843 & 12 58 29.6  & $+35$ 28 43 & QSO  & 1.92 & 0.1 & Y & N & 9 \\ 
J$1405.4+2223$& RIXOS F274-008 & 14 05 28.3  & $+22$ 23 33 & Sy1  & 0.156 & 0.2 & N & N & 10 \\ 
J$1426.8+2619$& Zw 1424.6+2632   & 14 26 50.0  & $+26$ 18 34 & Cl  & \nodata  & 1.0 & N & N & 11 \\  
J$1428.1+2337$& KUG 1425+238 & 14 28 07.7  & $+23$ 37 23 & G    & \nodata & 0.1 & N & N & 12  \\ 
J$1429.7+4240$& CRSS J1429.7+4240 & 14 29 45.1  & $+42$ 40 54 & QSO  & 1.67 & 0.2 & N & Y & 13 \\ 
J$1617.2+3454$& NGC6107 & 16 17 20.1  & $+34$ 54 05 & G/Group? & 0.03 & 0.1 & Y & N & 14 \\  
J$1728.2+5013$& 1 Zw 187 & 17 28 13.9  & $+50$ 13 10 & QSO  & 0.055 & 0.1 & Y & N & 15 \\  
J$1746.8+6836$& VII Zw742 & 17 47 00.1  & $+68$ 36 37 & G/Pair  & 0.063 & 0.2 & N & Y & 3  \\
J$1749.8+6823$& KUG 1750+683A & 17 49 50.6  & $+68$ 23 10 & Sy1  & 0.051 & 0.3 & Y & Y & 7 \\
J$1804.5+6938$& RIXOS F272-018 & 18 04 34.3  & $+69$ 37 37 & QSO  & 0.604 & 0.1 & N & N & 10 \\
J$1808.0+6948$& RIXOS F272-023 & 18 08 13.0  & $+69$ 48 06 & Sy1.8 & 0.096 & 0.1 & N & N & 11 \\
\tablecomments{
Col.(1): ASCA name;
Col.(2): NED name;
Col.(3): NED RA (2000);
Col.(4): NED DEC (2000);
Col.(5): NED classification: Cl = Cluster, G=Galaxy, QSO=Quasi Stellar Object; Sy=Seyfert; 
Col.(6): redshift;
Col.(7): Offset from ROSAT position (see Table~\ref{tab:rosat}) in arcmin;
Col.(8): Known radio source?;
Col.(9): Known IRAS/2MASS source?;
Col.(10): Reference
}
\tablerefs{
1. Maddox et al (1990);
2. Foltz et al. (1993);
3. Veron-Cetty \& Veron (1996);
4. Wisotzki et al. (2000);
5. Newberg et al. (1999);
6. Cutri et al. (2000);
7.  Fiore et al. (1999);
8. Young et al. (1998);
9. Becker et al. (1995);
10. Mason et al. (2000);
11. Zwicky \& Herzog (1963);
12. Takase \&  Miyauchi-Isobe (1985); 
13. Boyle et al. (1997);
14. Falco et al. (1999);
15. Johnston et al. (1995)
}
\enddata
\end{deluxetable}

\clearpage

\begin{deluxetable}{llllrl}

\tablecolumns{6}
\tablecaption{Mean hardness ratios \label{tab:hr}}
\tablehead{
\colhead{Sample} & \colhead{$N_{\rm obj}$} & 
\colhead{Method} & \colhead{Ratio}  & 
\colhead{Value}  & \colhead{$\Gamma$} \\
\colhead{(1)} & \colhead{(2)} & 
\colhead{(3)} & \colhead{(4)} & 
\colhead{(5)} & \colhead{(6)} 
}
\startdata
Full       & 69 & Unweighted & HM  & $-0.23 \pm 0.05$ & $\Gamma=0.7\pm0.2$\\
Full       & 69 & Unweighted & HS  & $-0.14 \pm 0.08$ & $\Gamma=0.9\pm0.2$\\
Full       & 69 & Unweighted & HR1 & $ 0.28 \pm 0.05$ & $\Gamma=1.1\pm0.1$\\
Full       & 69 & Weighted   & HM  & $-0.63 \pm 0.01$ & $\Gamma=2.1\pm0.1$ \\
Full       & 69 & Stacked    & HM  & $-0.52 \pm 0.03$ & $\Gamma=1.6\pm0.1$\\
Full-1     & 68 & Unweighted & HM  & $-0.22 \pm 0.05$ & $\Gamma=0.7\pm0.2$ \\
Full-1     & 68 & Weighted   & HM  & $-0.56 \pm 0.01$ & $\Gamma=1.8\pm0.1$ \\
Full-1     & 68 & Stacked    & HM  & $-0.46 \pm 0.03$ & $\Gamma=1.4\pm0.1$\\
Full-4 	   & 65 & Unweighted & HM  & $-0.21 \pm 0.05$ & $\Gamma=0.7\pm0.2$ \\
Full-4 	   & 65 & Weighted   & HM  & $-0.54 \pm 0.02$ & $\Gamma=1.7\pm0.1$\\
Full-4 	   & 65 & Stacked    & HM  & $-0.42 \pm 0.03$ & $\Gamma=1.3\pm0.1$\\ 
Defined HM & 53 & Unweighted & HM  & $-0.40 \pm 0.04$ & $\Gamma=1.3\pm0.2$ \\
$>6\sigma$ & 34 & Unweighted & HM  & $-0.43 \pm 0.05$ & $\Gamma=1.3\pm0.2$ \\
$>6\sigma$ & 34 & Unweighted & HS  & $-0.39 \pm 0.07$ & $\Gamma=1.3\pm0.1$\\
$>6\sigma$ & 34 & Unweighted & HR1 & $ 0.17 \pm 0.05$ & $\Gamma=1.3\pm0.1$\\
ROSAT      & 35 & Unweighted & HM  & $-0.44 \pm 0.05$ & $\Gamma=1.4\pm0.1$ \\
ROSAT      & 35 & Unweighted & HS  & $-0.44 \pm 0.07$ & $\Gamma=1.4\pm0.1$\\
Non-ROSAT  & 21 & Unweighted & HM  & $-0.01 \pm 0.08$ & $\Gamma=0.1\pm0.2$ \\
Non-ROSAT  & 21 & Unweighted & HS  & $ 0.14 \pm 0.07$ & $\Gamma=0.6\pm0.1$ \\
\tablecomments{
Col.(1): Sample or subsample. Full is the entire sample. Full-1 excludes
the brightest SHEEP source; Full-4 excludes the brightest four; Defined HM has
only the objects where the HM hardness ratio is constrained; $>6\sigma$
contains only sources above that significance level; ROSAT is the
ROSAT-detected objects; Non-ROSAT is the objects observed, but not
detected by ROSAT in pointed observations. 
Col.(2): Number of objects in subsample;
Col.(3): Method of calculation of mean hardness ratio (see text); 
Col.(4): Hardness ratio quoted HM is 5-10 vs 2-5 keV band; HS is 5-10 versus
0.6-2 keV band; HR1 is 2-10 vs 0.7-2 keV band (see text);
Col.(5): Hardness ratio value;
Col.(6): Equivalent photon spectral index
}
\enddata
\end{deluxetable}
\clearpage

\begin{deluxetable}{lllllccl}

\tablecolumns{7}
\tablecaption{Spectral fits to sources with S/N$>10$ \label{tab:fits}}

\tablehead{
\colhead{AX}          & 
\colhead{$N_{\rm H}$} & \colhead{$\Gamma$}      & 
\colhead{$F_{\rm X}(2-10)$} & \colhead{$F_{\rm X}(5-10)$} & 
\colhead{\chisq/d.o.f} & \colhead{ID} \\
\colhead{(1)} & \colhead{(2)} & 
\colhead{(3)} & \colhead{(4)} & 
\colhead{(5)} & \colhead{(6)} &
\colhead{(7)}
}
\startdata
J$0443.3-2820$ & $ 0.0^{+0.5}_{-0.0 } $ & $ 1.96^{+ 0.08}_{-0.08} $ &  1.97 & 0.87 & 166.1/176 & HE 0441-2826 \\
J$1040.3+2045$ & $ 0.0^{+3.0}_{-0.0 } $ & $ 1.69^{+ 0.25}_{-0.12} $ &  1.13 & 0.56 & 76.0/78 & \nodata \\
J$1115.3+4043$ & $ 0.9^{+1.7}_{-0.9 } $ & $ 1.97^{+ 0.19}_{-0.16} $ &  1.26 & 0.56 & 171.8/146 & 2MASS  \\
J$1220.2+0641$ & $ 0.0^{+3.0}_{-0.0 } $ & $ 1.74^{+ 0.33}_{-0.17} $ &  1.09 & 0.53 & 40.3/54 & \nodata \\
J$1531.9+2420$ & $ 0.0^{+2.4}_{-0.0 } $ & $ 1.80^{+ 0.28}_{-0.17} $ &  0.72 & 0.34 & 64.7/77 & FBQS \\
J$1728.2+5013$ & $ 0.0^{+0.4}_{-0.0 } $ & $ 2.32^{+ 0.06}_{-0.04} $ &  6.39 & 2.36 & 330.0/401 & I Zw 187 \\
J$1746.8+6836$ & $ 0.0^{+0.5}_{-0.0 } $ & $ 2.04^{+ 0.10}_{-0.10} $ &  1.69 & 0.72 & 136.3/129 & VII Zw 742 \\
J$2020.3-2226$ & $ 2.2^{+1.7}_{-1.6 } $ & $ 1.98^{+ 0.17}_{-0.17} $ &  2.40 & 1.06 & 119.3/111 & \nodata \\
\tablecomments{
Col.(1): ASCA name;
Col.(2): Absorbing colum density in units of $10^{21}$~cm$^{-2}$;
Col.(3): Power law photon index;
Col.(4): 2-10 keV flux in units of $10^{-12}$~erg cm$^{-2}$ s$^{-1}$;
Col.(5): 5-10 keV flux in units of $10^{-12}$~erg cm$^{-2}$ s$^{-1}$;
Col.(6): Fit statistic and degrees of freedom;
Col.(7): Optical identification from Table~\ref{tab:catids}, where
appropriate.}
\enddata
\end{deluxetable}


\end{document}